\documentclass[prb,twocolumn,preprintnumbers,amsmath,amssymb,nofootinbib,floatfix]{revtex4}

\usepackage{graphicx,bm,xcolor, bbold}
\usepackage[normalem]{ulem}
\usepackage{hyperref}

\makeatletter
\def\graphicscale{\twocolumn@sw{0.3}{0.4}}
\def\graphicthreescale{\twocolumn@sw{0.3}{0.4}}

\begin{document}

\title{Out-of-equilibrium finite-size scaling in generalized Kibble-Zurek
  protocols crossing  \\ quantum phase transitions in the presence of 
  symmetry-breaking perturbations}

\author{Francesca De Franco}
\affiliation{Dipartimento di Fisica dell'Universit\`a di Pisa,
        Largo Pontecorvo 3, I-56127 Pisa, Italy}

\author{Ettore Vicari} 
\affiliation{Dipartimento di Fisica dell'Universit\`a di Pisa
        and INFN, Largo Pontecorvo 3, I-56127 Pisa, Italy}

\date{\today}

\begin{abstract}
  We study the effects of symmetry-breaking perturbations in the
  out-of-equilibrium quantum dynamics of many-body systems slowly
  driven across a continuous quantum transition (CQT) by a
  time-dependent symmetry-preserving parameter. For this purpose, we
  analyze the out-of-equilibrium dynamics arising from generalized
  Kibble-Zurek (KZ) protocols, within a dynamic renormalization-group
  framework allowing for finite-size systems. We show that the time
  dependence of generic observables develops an out-of-equilibrium
  finite-size scaling (FSS) behavior, arising from the interplay
  between the time scale $t_s$ of the parameter variations in the KZ
  protocol, the size $L$ of the system, and the strength $h$ of the
  symmetry-breaking perturbation, in the limit of large $t_s$ and
  $L$. Moreover, scaling arguments based on the first-order adiabatic
  approximation of slow variations in quantum systems allow us to
  characterize the approach to the adiabatic regimes for some limits
  of the model parameters (for example when we take $t_s\to \infty$
  before $L\to\infty$), predicting asymptotic power-law suppressions
  of the nonadiabatic behaviors in the adiabatic limits.  This
  out-of-equilibrium FSS is supported by numerical analyses for the
  paradigmatic quantum Ising chain along generalized KZ protocols,
  with a time-dependent transverse field crossing its CQT, in the
  presence of a static longitudinal field breaking the ${\mathbb Z}_2$
  symmetry.
\end{abstract}

\maketitle


\section{Introduction}
\label{intro}

Out-of-equilibrium phenomena are generally observed in many-body
systems when they are driven across phase transitions, where
large-scale modes are not able to equilibrate even in the limit of
very slow changes of the system parameters. We mention hysteresis,
aging, out-of-equilibrium defect production, etc, which have been
addressed both theoretically and experimentally, at classical and
quantum phase transitions (see, e.g.,
Refs.~\onlinecite{Kibble-80,Binder-87,Bray-94,Zurek-96,CG-05,BDS-06,
  Dziarmaga-10,PSSV-11,Biroli-16,RV-21, Kibble-76, CDTY-91, BCSS-94,
  BBFGP-96, WNSBD-08,Ulm-etal-13, Pyka-etal-13, NGSH-15,
  Trenkwalder-etal-16} and references therein).  Many-body systems
develop out-of-equilibrium scaling behaviors when slowly crossing a
phase transition, in the limit of a large time scale $t_s$ of the
parameter variations driving the dynamics.  They generally depend on
the nature of the transition, whether it is driven by thermal or
quantum fluctuations, whether it is first-order or continuous, and in
the latter case on some global properties determining the universality
class of the critical behavior, see
e.g. Refs.~\onlinecite{Kibble-80,Zurek-85,Zurek-96,Polkovnikov-05, ZDZ-05,
  Dziarmaga-05, FFO-07,PG-08, Dziarmaga-10, DGP-10, GZHF-10, PSSV-11,
  CEGS-12, Dutta-etal-book,PV-16,PV-17,PRV-18-loc, PRV-18, RV-19-de,
  RDZ-19, RV-20,PRV-20,RV-21,RMAKVE-21,TV-22}.

In this paper we analyze the out-of-equilibrium scaling behaviors
emerging when one time-dependent symmetry-preserving parameter drives
a quantum many-body system across a continuous quantum transition
(CQT) in the presence of a time-independent symmetry-breaking
perturbation. For this purpose we consider generalized Kibble-Zurek
(KZ) protocols~\cite{RDZ-19}, where the out-of-equilibrium dynamics
arises from variations of one symmetry-preserving model parameter
$w(t)$ (defined so that $w_c=0$ is its critical value) across its
critical point, with a linear time dependence $w(t)=t/t_s$ and a large
time scale $t_s$, starting from the ground state associated with the
initial value $w(t_i)$. Analogous protocols have been generally
considered to discuss the so-called KZ problem related to the defect
production when crossing continuous transitions from disorder to order
phases, see e.g. Refs.~\onlinecite{Kibble-76, Kibble-80, Zurek-85, Zurek-96,
  ZDZ-05, Polkovnikov-05, Dziarmaga-05, PG-08, Dziarmaga-10,
  Dutta-etal-book, CEGS-12, PSSV-11, Damski-05, USF-07, USF-10,
  NDP-13, RDZ-19, RMAKVE-21}.

We investigate this issue within renormalization-group (RG) scaling
frameworks at CQTs~\cite{SGCS-97,Sachdev-book,CPV-14,RV-21}, which
allow us to develop an out-of-equilibrium finite-size scaling (FSS)
theory~\cite{RV-21,RV-20,TV-22} describing the intricate interplay
among the Hamiltonian parameters, the time scale $t_s$ of the KZ
protocol and the lattice size $L$, in the limits of large $t_s$ and
large $L$, allowing for the effects of a further (sufficiently small)
symmetry-breaking perturbation.  The effects of symmetry-breaking
perturbations within KZ protocols have been also addressed in
Ref.~\onlinecite{RDZ-19}, where the scaling behaviors in the infinite-volume
(thermodynamic) limit were analyzed.  Here we extend the
characterization of the out-of-equilibrium scaling behavior to finite
systems, in an appropriate out-of-equilibrium FSS limit.

As a paradigmatic model where to test the out-of-equilibrium FSS
framework, we consider the quantum Ising chain.  We analyze the
out-of-equilibrium FSS associated with generalized KZ protocols, and
provide numerical results that support it.  In particular, we analyze
the approach to the adiabatic regimes that can be realized for some
limits of the parameters, involving the time scale $t_s$, the size of
the system, and the strength of the symmetry-breaking perturbation.
Within the same theoretical framework, we also discuss more general KZ
protocols in which both symmetry-preserving and symmetry-breaking
terms are time dependent, and driven across the transition point.

We remark that equilibrium and out-of-equilibrium FSS frameworks
generally simplify the study of the universal features of critical
behaviors.  This is essentially related to the fact that the general
requirement of a large length scale $\xi$ of the critical correlations
is not subject to further conditions on the system size $L$. Indeed
$\xi\sim L$ for FSS, while critical behaviors in the thermodynamic
limit requires $\xi\ll L$. Therefore much larger systems are necessary
to probe analogous length scales $\xi$ in the thermodynamic limit.
The FSS scenarios are often observed for systems of moderately large
size, see e.g. Refs.~\onlinecite{PRV-18,RV-20,RV-21,TV-22}.  Therefore, FSS
behaviors may be more easily accessed by experiments where the
coherent quantum dynamics of only a limited number of particles or
spins can be effectively realized, such as experiments with quantum
simulators in laboratories, e.g., by means of trapped
ions~\cite{Islam-etal-11, Debnath-etal-16}, ultracold
atoms~\cite{Simon-etal-11, Labuhn-etal-16}, or superconducting
qubits~\cite{Salathe-etal-15, Cervera-18}.

The paper is organized as follows. In Sec.~\ref{modprot} we describe
generalized KZ protocols in the presence of a further static
symmetry-breaking perturbation. We also present the paradigmatic
$d$-dimensional quantum Ising models, and in particular quantum Ising
chains, which provide a theoretical laboratory to address the
out-of-equilibrium dynamics along generalized KZ protocols.  In
Sec.~\ref{qfssoneway} we derive the out-of-equilibrium FSS laws
emerging along the generalized KZ protocols in the presence of a
static symmetry-breaking perturbation, which are supposed to apply to
generic CQTs.  This scaling framework is supported by numerical
results for the quantum Ising chain, based on exact diagonalization
up to size $L\approx 20$.  In Sec.~\ref{adiabatic} we discuss the
approach to adiabatic regimes, which are always possible in
finite-size systems, and/or in the presence of a further external
relevant perturbation, for a sufficiently large time scale $t_s$ of
the parameter variations. Sec.~\ref{bothtimedep} presents a brief
discussion of the more general case in which the KZ protocol is
further extended to the case in which both the symmetry-preserving and
the symmetry-breaking parameters are time dependent.  Finally, in
Sec.~\ref{conclu} we summarize and draw our conclusions.

\section{The models and dynamic protocols}
\label{modprot}

\subsection{Kibble-Zurek protocols}
\label{KZprot}

We consider quantum many-body systems whose Hamiltonian can be written
as
\begin{equation}
\hat H(w,h) = \hat H_{c} + w \, \hat H_{w} + h \hat H_h \,,
  \label{hlamt}
\end{equation}
where $\hat H_c$ is a critical Hamiltonian (i.e. with its parameters
tuned to their critical values), $w$ is associated with a relevant
perturbation $\hat H_w$ preserving the symmetry, and the parameter $h$
is another relevant parameter associated with a symmetry-breaking term
$\hat H_h$.  Both Hamiltonian parameters $w$ and $h$ vanish at the
critical point, i.e. $w_c=h_c=0$.  We assume that for $h=0$ the
critical point $w_c=0$ separates disordered ($w<0$) and ordered
($w>0$) phases. On the other hand, nonzero values of the parameter $h$
gives always rise to a gapped phase.

For sufficiently small values of $h$ (we will make this condition more
precise below), quasi-adiabatic passages through the continuous
quantum transition are obtained by slowly varying $w$ across $w_c =
0$, following, e.g., the KZ protocol:

(i) The quantum evolution of finite systems of size $L$ starts at the
time $t_i$ from the ground state
$|\Psi_0(w_i,h)\rangle$
associated with the initial value $w_i<0$
corresponding to the disordered phase.
  
(ii) Then the system evolves unitarily according to the Schr\"odinger
equation
  \begin{eqnarray}
&&    {d \, |\Psi(t)\rangle \over d t} =
    - i \, \hat H[w(t),h] \, |\Psi(t)\rangle \,,
    \label{unitdyn}\\
&&|\Psi(t=0)\rangle = |\Psi_0(w_i,h)\rangle\,,
\nonumber
  \end{eqnarray}
with a linear time dependence of $w(t)$,
  \begin{equation}
    w(t) = t/t_s \,,
    \label{wtkz}
  \end{equation}
  up to a final value $w_f>0$, corresponding to parameter values
  within the ordered phase. Thus we have $t_i = t_s \, w_i<0$ and
  $t_f= t_s \, w_f>0$.  The parameter $t_s$ of the KZ protocol
  represents the time scale of the time dependence of the Hamiltonian
  parameter $w$,

  Across a phase transition, in particular in the absence of further
  relevant perturbations, i.e. $h=0$, the growth of an
  out-of-equilibrium dynamics is inevitable in the thermodynamic limit
  (i.e. $L\to\infty$ before taking the critical limit) even for very
  slow changes of the parameter $w$, because large-scale modes are
  unable to equilibrate the long-distance critical correlations
  emerging at the transition point. As a consequence, when starting
  from equilibrium states at the initial value $w_i$, the system
  cannot pass through equilibrium states associated with the values of
  $w(t)$ across the transition point, thus departing from an adiabatic
  dynamics. Such a departure from equilibrium develops peculiar
  out-of-equilibrium scaling phenomena in the limit of large time
  scale $t_s$ of the time variation of $w(t)$.

  This out-of-equilibrium scenario substantially changes in the
  presence of a finite nonzero symmetry-breaking perturbation,
  i.e. Hamiltonian (\ref{hlamt}) with $h\neq 0$.  This is essentially
  due to the fact that the gap $\Delta$ does not generally close when
  $h\neq 0$.  Indeed, at the critical point $w=w_c$, $\Delta$ remains
  finite in the large-$L$ limit, behaving as $\Delta \sim
  |h|^{\varepsilon}$ ($\varepsilon>0$) for small $h$ (standard RG
  arguments show that $\varepsilon=z/y_h$ where $z$ and $y_h$ are
  appropriate critical exponents associated with the universality
  class of the CQT,~\cite{Sachdev-book} see below). Therefore, the
  adiabatic evolution through the ground states associated with the
  instantaneous values of $w(t)$ can be always realized for
  sufficiently large time scale $t_s$. In the presence of the
  symmetry-breaking perturbation $h$, a nontrivial out-of-equilibrium
  FSS limit can be still defined by appropriately rescaling $h$ in the
  large-$L$ limit, providing information for sufficiently small values
  of $h$. This issue will be addressed in the next sections within a
  general out-of-equilibrium FSS framework for quantum many-body
  systems driven across CQTs, see e.g. Ref.~\onlinecite{RV-21}.

It is worth mentioning that a related issue is the so-called KZ
problem, i.e.  the scaling behavior of the amount of final defects
after slow passages through continuous transitions, from the disorder
phase to the order phase~\cite{Kibble-76, Kibble-80, Zurek-85,
  Zurek-96, ZDZ-05, Polkovnikov-05, Dziarmaga-05, PG-08, Dziarmaga-10,
  Dutta-etal-book, PSSV-11, CEGS-12, RV-21, TV-22, Damski-05, USF-07,
  USF-10, NDP-13, RMAKVE-21}. The general features of the KZ scaling,
and in particular the KZ predictions for the abundance of residual
defects, have been confirmed by several analytical and numerical
studies, see, e.g., Refs.~\onlinecite{Dziarmaga-10, Dutta-etal-book,
  PSSV-11, CEGS-12, NDP-13, RV-21} and citing references, and by
experiments for various physically interesting systems, see, e.g.,
Refs.~\onlinecite{DRGA-99, MMARK-06, SHLVS-06, WNSBD-08, CWBD-11,
  Griffin-etal-12, Ulm-etal-13, Pyka-etal-13, LDSDF-13, Braun-etal-15,
  Chomaz-etal-15, NGSH-15, Cui-etal-16, Gong-etal-16, Anquez-etal-16,
  CFC-16, Keesling-etal-19}.

\subsection{The quantum Ising models}
\label{qischain}

As paradigmatic quantum many-body systems we consider the
$d$-dimensional quantum Ising models in the presence of transverse and
longitudinal fields, described by the Hamiltonian
\begin{eqnarray}
  \hat{H}(g,h) = - J \sum_{\langle {\bm x} {\bm y}\rangle}
  \hat\sigma^{(1)}_{{\bm x}\phantom{1}}
  \hat\sigma^{(1)}_{\bm y} - g \sum_{{\bm x}} \hat\sigma^{(3)}_{\bm x}
  - h \sum_{\bm x} \hat\sigma^{(1)}_{\bm x}\,\quad
  \label{qisingmodel}
\end{eqnarray}
defined on a cubic-like lattice, where $\hat\sigma^{(k)}_{\bm x}$ are
the Pauli matrices on the site ${\bm x}$ ($k = 1,2,3$ labels the three
spatial directions), and the first sum runs on the bonds ${\langle
  {\bm x} {\bm y}\rangle}$ of the lattice. In the following we
consider quantum Ising systems of size $L$ with periodic boundary
conditions (PBC).  We assume ferromagnetic
nearest-neighbour interactions with $J=1$.  We set the Planck constant
$\hslash=1$.

Transforming the spin operators as $\hat\sigma_{\bm x}^{(1)}\to
-\hat\sigma_{\bm x}^{(1)}$ and $\hat\sigma_{\bm x}^{(3)}\to
\hat\sigma_{\bm x}^{(3)}$, the Hamiltonian $\hat H(g,h)$ maps into
$\hat H(g,-h)$, thus $\hat H$ is ${\mathbb Z}_2$ symmetric for $h=0$.
The quantum Ising models are always gapped in the presence of a
nonzero symmetry-breaking term, i.e. for $h\neq 0$.  They undergo a
CQT at $g=g_c$ and $h=0$, whose quantum critical behavior belongs to
the $(d+1)$-dimensional Ising universality class, due to the
quantum-to-classical mapping, see
e.g. Refs.~\onlinecite{Sachdev-book,RV-21}.  The parameters $w\equiv g_c-g$
and $h$ are relevant at the CQT, being respectively associated with
the leading even and odd RG perturbations at the $(d+1)$-dimensional
Ising fixed point.

Several exact results are known for one-dimensional models, such as
the location of the critical point, at $g_c=1$, and the RG dimensions
of the Hamiltonian parameters $w$ and $h$, which are $y_w=1/\nu=1$ and
$y_h=15/8$ respectively.  Accurate estimates are available for
two-dimensional quantum Ising systems, see, e.g.,
Refs.~\onlinecite{PV-02,GZ-98,CPRV-02,Hasenbusch-10,KPSV-16,KP-17,Hasenbusch-21},
in particular Ref.~\onlinecite{KPSV-16} reports $y_w=
1/\nu=1.58737(1)$ and $y_h = 2.481852(1)$.  For $d=3$, the critical
exponents take their mean-field values, $y_w=2$ and $y_h=3$, however
the critical singular behavior presents additional multiplicative
logarithmic factors~\cite{Sachdev-book,RV-21,PV-02}.  The length scale
$\xi$ of the critical modes behaves as $\xi\sim |w|^{-\nu}$ for $h=0$,
and $\xi\sim |h|^{-1/y_h}$ at $w=w_c=0$.  The dynamic exponent $z$,
controlling the vanishing $\Delta\sim \xi^{-z}$ of the gap at the
transition point, is given by $z=1$ in any dimension.  We also recall
that the RG dimension of the order-parameter field, associated with
the longitudinal operators $\hat\sigma_{\bm x}^{(1)}$, is given by
\begin{equation}
  y_l=d + z - y_h\,,
  \label{ylyh}
\end{equation}
while that associated with the transverse operators $\hat\sigma_{\bm
  x}^{(3)}$ is given by $y_t = d + z - y_w $.

The KZ protocols outlined in Sec.~\ref{KZprot} can be implemented
within quantum Ising chains, by identifying the terms of the generic
Hamiltonian (\ref{hlamt}) with
\begin{eqnarray}
  &&  \hat H_c =
- \sum_{\langle {\bm x} {\bm y}\rangle}
  \hat\sigma^{(1)}_{{\bm x}\phantom{1}}
  \hat\sigma^{(1)}_{\bm y} - g_c \sum_{{\bm x}} \hat\sigma^{(3)}_{\bm x}\,,
  \label{hcdef}\\
  && w(t) = g_c-g(t)\,,\;\; \hat H_w =  \sum_{\bm x}
  \hat\sigma^{(3)}_{\bm x}\,,\;\;
\hat H_h =  -\sum_{\bm x} \hat\sigma^{(1)}_{\bm x}\,.
\nonumber
\end{eqnarray}

\subsection{Observables monitoring the quantum evolution}
\label{obs}
  
The out-of-equilibrium evolution of quantum many-body systems
resulting from the KZ protocol can be monitored looking at the
behavior of some observables and correlations at fixed time.

To characterize the departure from adiabaticity along the slow dynamic
across the CQT, we monitor the adiabaticity function
\begin{eqnarray}
  A(t) = |\langle \, \Psi_0[w(t),h] \, | \, \Psi(t) \, \rangle|\,,
  \label{adtfunc}
\end{eqnarray}
where $|\,\Psi_0[w(t),h]\,\rangle$ is the ground state of the
Hamiltonian $\hat H[w(t),h]$, i.e. at the instantaneous value $w(t)$,
while $|\,\Psi(t)\,\rangle$ is the actual time-dependent state
evolving according to the Schr\"odinger equation (\ref{unitdyn}).  The
adiabaticity function measures the overlap of the time-dependent state
at a time $t$ with the ground state of the Hamiltonian at the
corresponding $w(t)$.  Since the KZ protocol starts from the ground
state associated with $w_i=w(t_i)$, we have $A(t_i) = 1$ initially. Of
course, the adiabaticity function for an adiabatic evolution takes the
value $A(t)=1$ at any time $t>t_i$.

In general protocols crossing transition points, $A(t)$ is expected to
depart from the initial value $A(t_i)=1$, due to the impossibility of
the system to adiabatically follow the changes of the function $w(t)$
across its critical value $w=0$. Note however that this is strictly
true in the infinite-volume limit.  In systems of finite size $L$,
there is always a sufficiently large time scale $t_s$, so that the
system can evolve adiabatically, essentially because finite-size
systems are strictly gapped, although the gap $\Delta$ at the
continuous quantum transition gets suppressed as $\Delta \sim
L^{-z}$. The interplay between the size $L$ and the time scale $t_s$
gives rise to nontrivial out-of-equilibrium scaling behaviors, which
can be studied within out-of-equilibrium FSS
frameworks~\cite{RV-21,RV-20,TV-22}.

Another global observable monitoring the departure from adiabaticity
is provided by the surplus energy of the system with respect to its
instantaneous ground state at $w(t)$, often called {\em excitation}
energy in earlier works, i.e.
\begin{equation}
  E_s(t) = \langle \Psi(t) | \, \hat{H} \, | \Psi(t) \rangle - \langle
  \Psi_0[w(t)] | \, \hat{H} \,| \Psi_0[w(t)] \rangle \,.
  \label{etdiff}
  \end{equation}
Since the KZ protocols that we consider start from a ground state at
$t_i$, the excitation energy $E_s(t)$ vanishes along adiabatic
evolutions, while nonzero values $E_s(t)>0$ are related to the degree
of out-of-equilibrium of the dynamics across the transition.  One may
also consider the corresponding density of excitation energy
\begin{equation}
  D_e(t) = L^{-d} E_s(t)\,.
  \label{dedef}
  \end{equation}

To monitor the out-of-equilibrium dynamics of the spin expectation
values and correlations, we consider the evolution of the local
and global average magnetization
\begin{equation}
  m_{\bm x}(t) \equiv \langle \Psi(t) | \, \hat\sigma_{\bm x}^{(1)} \,
  | \Psi(t)\rangle\,,
  \;\;\; M(t) \equiv {1\over L^{d}} \sum_{\bm x} m_{\bm x}(t)\,,
  \label{magnt}
\end{equation}
the fixed-time correlation function 
\begin{equation}
  G(t,{\bm x},{\bm y}) \equiv \langle \Psi(t) | \, \hat\sigma_{\bm x}^{(1)} \,
  \hat\sigma_{\bm y}^{(1)}\,| \Psi(t)\rangle\,.
  \label{twopointt}
\end{equation}
In the absence of boundaries, such as the case of PBC, translation
invariance implies $m_{\bm x}(t) = M(t)$ and $G(t,{\bm x},{\bm y}) \equiv
G(t,{\bm x}-{\bm y})$.

In the following we outline the out-of-equilibrium FSS scenario
applying to the generalized KZ protocol described in
Sec.~\ref{KZprot}.  We contextualize it within quantum Ising models.
However, most scaling arguments can be straightforwardly extended to
generic CQTs.  To support the emerging out-of-equilibrium FSS
behaviors, we also report numerical analyses for the one-dimensional
Ising chain with a time-dependent transverse field $g(t)$ and a static
longitudinal field $h$.  They are based on exact-diagonalization
methods. The corresponding Schr\"odinger equation is solved using a
$4^{\rm th}$ order Runge-Kutta method. This approach allows us to
compute the out-of-equilibrium dynamics for lattice size up to
$L\approx 20$, with high accuracy (practically exact).  As we shall
see, such moderately large (or relatively small) systems turn out to
be already sufficient to provide a robust evidence of the
out-of-equilibrium FSS outlined in the paper.

\section{Out-of-equilibrium scaling}
\label{qfssoneway}

We discuss here the out-of-equilibrium scaling behaviors emerging
along the KZ protocol outlined in Sec.~\ref{KZprot}, within a dynamic
RG framework~\cite{RV-21}.  Out-of-equilibrium FSS laws are expected
to develop in the limit of a large time scale $t_s$ of the driving
parameter $w(t)$, large size $L$ of the system, and for sufficiently
small values of the symmetry-breaking parameter $h$. They describe the
interplay of the various scales of the problem, such as the time $t$
and time scale $t_s$ of the KZ protocol, the size $L$ of the system,
the energy scale $\Delta \sim L^{-z}$ of the system at the critical
point, and the external longitudinal field $h$.

\subsection{Homogenous scaling laws}
\label{homscalaws}

Let us consider observables constructed from a local operator
$\hat{O}({\bm x})$. The general working hypothesis underlying
out-of-equilibrium FSS frameworks is that the expectation value of
$\hat{O}({\bm x})$ and its correlation functions along KZ protocols
obey asymptotic homogeneous scaling laws,~\cite{RV-21} such as
  \begin{eqnarray}
&& O(t,t_s,w_i,h,L) \equiv\langle \Psi(t)| \hat{O}({\bm x})
    |\Psi(t)\rangle \label{Oscadynwt} \\ && \;\;\approx b^{-y_o} {\cal
      O}(b^{-z} t, b^{y_w} w_i, b^{y_w} w(t),  b^{y_h} h,L/b)\,,
    \nonumber
    \\ 
    &&    G_O({\bm x},t,t_s,w_i,h,L) \equiv
\langle \Psi(t)| \hat{O}({\bm x}_1)\,\hat{O}({\bm x}_2)
  |\Psi(t)\rangle
    \label{g12scadynwt} \\
&& \;\;\approx
    b^{-2y_o} \, {\cal G}({\bm x}/b, b^{-z} t,
    b^{y_w} w_i, b^{y_w} w(t), b^{y_h} h, L/b)\,,
    \qquad \nonumber
  \end{eqnarray}
  where $b$ is an arbitrary (large) length scale, the RG dimension
  $y_o$ of the local operator $\hat{O}$ and the RG exponents $y_w$,
  $y_h$, and $z$, are determined by the universality class of the
  CQT. We assumed translation invariance, i.e., systems without
  boundaries such as those with PBC, so that the expectation value $O$
  does not depend on ${\bm x}$, and the two-point function depends on
  the difference ${\bm x}\equiv {\bm x}_1 - {\bm x}_2$ only.
  
  Note that in the above homogenous scaling laws the dynamic features
  are essentially encoded in the time dependence of the scaling
  functions, and in particular through the time-dependent Hamiltonian
  parameter $w(t)$. The other features are analogous to those arising
  from equilibrium FSS at CQTs~\cite{CPV-14,RV-21}, where the arguments
  of the scaling functions take into account the RG dimensions $y_w$
  and $y_h$ of the relevant parameters $w$ and $h$ at the RG fixed
  point associated with the CQT. In this respect the RG scaling
  framework for KZ protocols is obtained by replacing $w$ with $w(t)$
  in the equilibrium homogenous scaling laws.  The scaling functions
  ${\cal O}$ and ${\cal G}_O$ are expected to be universal, i.e.
  largely independent of the microscopic details of the models and the
  KZ protocols (apart from a multiplicative factor and normalizations
  of the arguments).

By taking the ratio between the arguments $b^{-z}t$ and $b^{y_w} w(t)$
of the scaling functions reported in Eqs.~(\ref{Oscadynwt}) and
(\ref{g12scadynwt}), we obtain the scaling variable $b^{-(y_w+z)}
t_s$, which tells us how the time scale $t_s$ must be rescaled to
observe the out-of-equilibrium scaling behavior. Then, by exploiting
the arbitrariness of $b$, we may set
\begin{equation}
    b^{-(y_w+z)} t_s=1\,,
    \label{bfix}
    \end{equation}
to derive the length scale
  \begin{equation}
  \lambda = t_s^{1/\zeta} \,,\qquad \zeta = y_w + z\,,
  \label{lambdats}
  \end{equation}
corresponding to the length scale arising from the linear time
dependence $w(t)=t/t_s$.

Out-of-equilibrium FSS can be straightforwardly derived by fixing $b=L$
in Eqs.~(\ref{Oscadynwt}) and (\ref{g12scadynwt}).  Then, the
asymptotic out-of-equilibrium FSS limit is obtained by taking
$t_s\to\infty$ and $L\to\infty$, while appropriate scaling variables
are kept fixed, such as~\cite{RV-21}
\begin{eqnarray}
  &\Upsilon = t_s/L^{\zeta} = (\lambda/L)^\zeta\,,\quad
    &K = w(t) L^{y_w}\,,
  \label{scalvars}\\
  &\Phi = h L^{y_h}\,,\qquad\qquad\quad\;\;
  &\Sigma = h \, \lambda^{y_h}\,,\nonumber\\
  &\Theta_i=w_i\, \lambda^{y_w} \,,\qquad\qquad\;\;
  &\Theta = w(t) \,\lambda^{y_w}
  = t / t_s^{\kappa} \,,
  \nonumber
\end{eqnarray}
where
\begin{equation}
0< \kappa=z/\zeta<1\,.
\label{kappadef}
\end{equation}
We obtain $\kappa=1/2$ for one-dimensional Ising chain, $\kappa
\approx 0.386$ for $d=2$, and $\kappa = 1/3$ for $d=3$.  Note that the
above scaling variables are not all independent. Indeed one can easily
check that $K= \Upsilon^{\kappa-1}\Theta$, $\Sigma = \Phi
\Upsilon^{y_h/\zeta}$, and $\Theta\ge \Theta_i$.  Moreover, the time
scaling variable $t \,\Delta$, where $\Delta \sim L^{-z}$ is the
critical gap of the system, can be straightforwardly related to
$\Theta$ and $\Upsilon$ by $t \, \Delta \sim \Theta \Upsilon^\kappa$.
On the other hand, the out-of-equilibrium FSS keeping $\Upsilon$ fixed
implies that the KZ time scale $t_s$ is generally much larger that the
inverse energy scale extracted from the energy differences of the
lowest levels at the critical point. Indeed, we have that
\begin{equation}
t_s \Delta(L) \sim t_s L^{-z} = \Upsilon L^{y_w} \to\infty\,.
\label{tsde}
\end{equation}

\subsection{Out-of-equilibrium FSS}
\label{FSS}

We can use the general homogenous scaling laws reported in
Eqs.~(\ref{Oscadynwt}) and (\ref{g12scadynwt}) to derive an
out-of-equilibrium FSS limit, and the behaviors of the observables of
the quantum Ising models in this limit, such as the longitudinal
magnetization $M$ and the correlation function $G$ of the quantum
Ising systems, the adiabaticity function and the excitation energy,
defined in Sec.~\ref{obs}.

By fixing $b=L$, we write their asymptotic behavior in the
out-of-equilibrium FSS limit in terms of the scaling variables
$\Upsilon$, $\Theta$, $\Phi$, and $\Theta_i$ defined in
Eqs.~(\ref{scalvars}), as~\cite{RV-21}
\begin{eqnarray}
  M(t,t_s,w_i,h,L) &\approx&
  L^{-y_l} {\cal M}_i(\Upsilon,\Theta,\Phi,\Theta_i)\,,
  \label{magsca}\\
  G({\bm x},t,t_s,w_i,h,L) &\approx&
  L^{-2y_l}\, {\cal G}_i({\bm x}/L,\Upsilon,\Theta,\Phi,\Theta_i)\,,\quad
  \label{G2sca}
\end{eqnarray}
where ${\cal M}_i$ and ${\cal G}_i$ are scaling functions, expected
to be largely  universal with respect to the details of
the model and KZ protocol.

Since the KZ protocol runs within the interval $t_i\le t \le t_f$,
corresponding to the interval $w_i\le w(t) \le w_f$, the scaling
variable $\Theta$ takes values within the interval
\begin{equation}
  \Theta_i \le \Theta \le \Theta_f \equiv w_f t_s^{1-\kappa}>0\,.
  \label{intomega}
  \end{equation}
We omit the dependence on $\Theta_f$, because the out-of-equilibrium
FSS limit at fixed $\Theta<\Theta_f$ does not depend on $\Theta_f$,
but only on $\Upsilon$ and $\Theta_i$.  Of course, if we keep $w_f$
fixed in the large-$t_s$ limit, i.e. if we do not scale $w_f$ to zero
to keep $\Theta_f$ fixed, then $\Theta_f\to\infty$.

With increasing $L$, the out-of-equilibrium FSS develops  within a
smaller and smaller interval $\delta w$ of values of $|w|$ around
$w=0$.  In particular, for the {\em most critical} case when $h=0$,
the time interval of the out-of-equilibrium process described by the
scaling laws scales as $t_{\rm KZ}\sim t_s^{\kappa}$, thus the
relevant interval $\delta w$ of values of $|w|$, where a nontrivial
out-of-equilibrium scaling behavior is observed, must shrink as
\begin{equation}
  \delta w \sim {t_{\rm KZ}/t_s}\sim L^{-y_w}\,,
  \label{deltaw}
\end{equation}
when keeping $\Upsilon$ fixed. Therefore, assuming that the KZ
protocol starts from a gapped phase, see Sec.~\ref{KZprot}, and that
the initial $w_i<0$ is kept fixed (corresponding to $\Theta_i\to
-\infty$), the same out-of-equilbrium FSS is expected to hold,
irrespective of the value of $w_i$.  Therefore, the out-of-equilibrium
FSS at fixed $w_i<0$ simplifies to
\begin{eqnarray}
  &&M(t,t_s,w_i,h,L) \approx L^{-y_l}
    {\cal M}(\Upsilon,\Theta,\Phi)\,,
  \label{magsca2}\\
    &&G({\bm x},t,t_s,w_i,h,L) \approx L^{-2y_l}\, {\cal
      G}({\bm x}/L,\Upsilon,\Theta,\Phi)\,,
  \label{G2sca2}
\end{eqnarray}
being independent of $w_i$.  The scaling functions ${\cal M}$ and
${\cal G}$ are expected to match the $\Theta_i\to -\infty$ limit of
the scaling functions depending on $\Theta_i$ (when $w_i$ gets
appropriately rescaled to keep $\Theta_i$ fixed),
cf. Eqs.~(\ref{magsca}) and (\ref{G2sca}). Thus,
\begin{equation}
{\cal M}(\Upsilon,\Theta,\Phi) = {\cal
  M}_i(\Upsilon,\Theta,\Phi,\Theta_i\to -\infty)\,,
\label{thetaiinf}
\end{equation}
and analogously for the correlation function $G$.

Since the magnetization vanishes for $h=0$ for symmetry reasons, we 
must have that
\begin{eqnarray}
{\cal M}(\Upsilon,\Theta,\Phi=0) = 0\,.
\label{zeromag}
\end{eqnarray}
Actually, in finite systems the magnetization, as well as any other
correlation and observables, is expected to be an analytical function
of $h$, thus $M\sim h$ at small $h$.  Therefore, when keeping the
other scaling argument fixed (in particular for $\Upsilon>0$, since
$\Upsilon\to 0$ corresponds to the thermodynamic limit), we should
have that
\begin{eqnarray}
  {\cal M}(\Upsilon,\Theta,\Phi) =
  c\,\Phi + O(\Phi^3)\,,
\label{zeromag2}
\end{eqnarray}
where the coefficient $c$ depends on the other scaling variables [an
  analogous behavior is expected at equilibrium, i.e.  ${\cal M}_{\rm
    eq}(K,\Phi) = c(K)\,\Phi + O(\Phi^3)$].  Note also that
$M(t,t_s,w_i,h,L)=-M(t,t_s,w_i,-h,L)$, thus
\begin{equation}
{\cal
  M}(\Upsilon,\Theta,\Phi)=-{\cal
  M}(\Upsilon,\Theta,-\Phi)\,.
\label{msymm}
\end{equation}

The out-of-equilibrium FSS relations can be also written in
alternative ways, using other equivalent sets of scaling variables (as
we shall see, some interesting limits, such as the thermodynamic and
adiabatic limits, are defined for particular choices of the scaling
variables).  For example, one may write them in terms of $K$ instead
of $\Theta$, such as
\begin{eqnarray}
  M(t,t_s,w_i,h,L) \approx
  L^{-y_l} \widetilde{\cal M}(\Upsilon,K,\Phi)\,,
\label{magscaK}
\end{eqnarray}
by replacing $\Theta = \Upsilon^{1-\kappa}K$
in Eq.~(\ref{magsca2}). As we shall see, the scaling variables
$\Upsilon$, $K$ and $\Phi$ are most appropriate to discuss the
adiabatic limit.

\begin{figure}[!htb]
  \includegraphics[width=0.95\columnwidth]{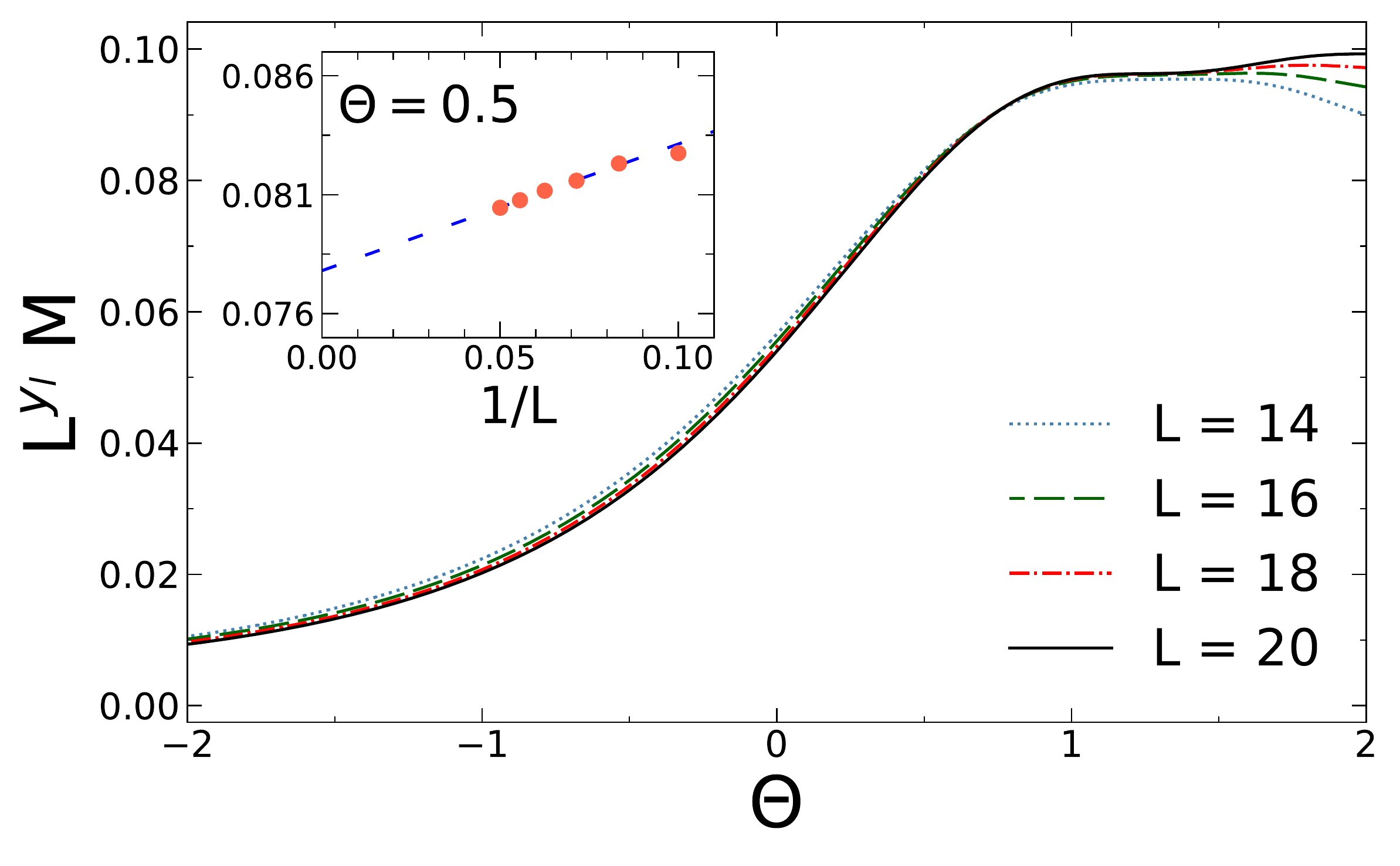}
  \includegraphics[width=0.95\columnwidth]{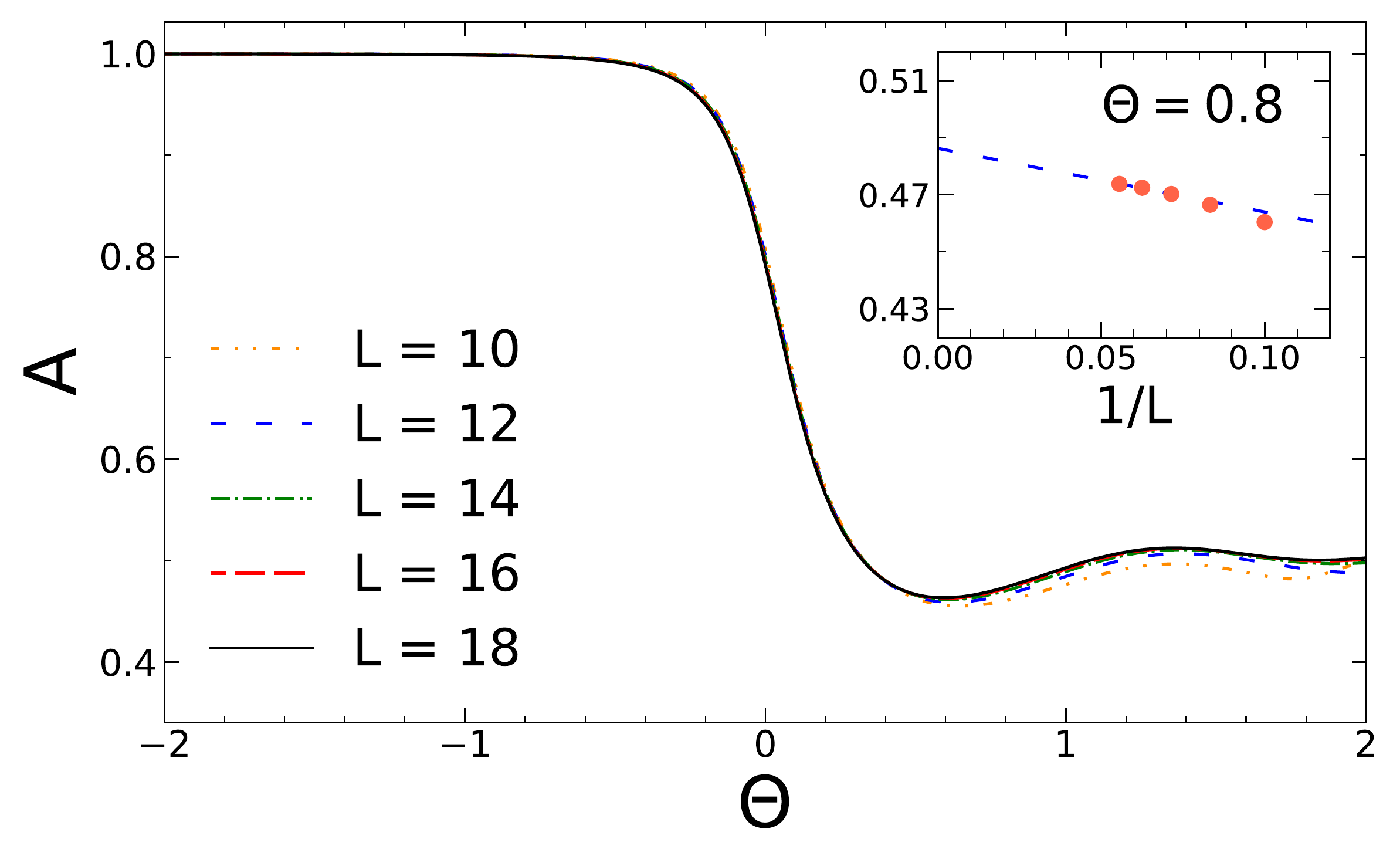}
  \caption{Some results for the time evolution of quantum Ising chains
    along the generalized KZ protocol outlined in Sec.~\ref{KZprot},
    keeping $w_i = -3$, $\Upsilon\equiv t_s/L^\zeta=0.01$ (we recall
    that $\zeta=2$), and $\Phi\equiv h L^{y_h}=1$ ($y_h=15/8$) fixed,
    versus $\Theta= t / t_s^{\kappa}$ with $\kappa=1/2$.  We show results
    for the magnetization (top), for which $y_l=1/8$, and the
    adiabaticity function (bottom), up to $L=20$ and $L=18$
    respectively.  Their behaviors nicely agree with the asymptotic
    out-of-equilibrium FSS reported in Eqs.~(\ref{magsca2}) and
    (\ref{wsca2}). The insets of both figures report data at fixed
    $\Theta$ versus $1/L$. They appear substantially consistent with
    an $1/L$ convergence to the asymptotic FSS.  }
  \label{dfss}
\end{figure}

An out-of-equilibrium FSS behavior analogous to that in
Eq.~(\ref{magsca2}) is put forward for the adiabaticity function,
cf. Eq.~(\ref{adtfunc}),
\begin{eqnarray}
  A(t,t_s,w_i,h,L) \approx {\cal A}(\Upsilon,\Theta,\Phi)
  = \widetilde{\cal A}(\Upsilon,K,\Phi)\,,\quad
\label{wsca2}
\end{eqnarray}
when keeping $w_i<0$ fixed.  Due to the initial condition of the KZ
protocol, cf. Eq.~(\ref{unitdyn}), i.e. the ground state at $w_i$, we
must have $A(t_i,t_s,w_i,h,L)=1$, and therefore ${\cal
  A}(\Upsilon,\Theta\to -\infty,\Phi)=1$.  It is therefore natural to
assume that its scaling behavior does not require a power of the size
as prefactor, unlike the magnetization, cf. Eq.~(\ref{magsca2}).
Using standard RG arguments, we may also derive an ansatz for the
out-of-equilibrium FSS behavior of the excitation energy defined in
Eq.~(\ref{etdiff}), which turns out to be
\begin{eqnarray}
  E_s(t,t_s,w_i,h,L) \approx L^{-z} {\cal E}(\Upsilon,\Theta,\Phi)
=L^{-z} \widetilde{\cal E}(\Upsilon,K,\Phi)\,,\quad
\label{essca}
\end{eqnarray}
where $z=1$ is the RG exponent associated with the energy differences
of the lowest states of the spectrum.  Note that the leading analytic
background contributions~\cite{CPV-14,RV-21}, generally arising at the
critical point, get cancelled by the difference of the two terms in
the definition of $E_s$, cf. Eq.~(\ref{etdiff}), thus justifying the
scaling ansatz (\ref{essca}), where the excitation energy is assumed
to scale as the energy gap at the transition point, i.e. $\Delta\sim
L^{-z}$.

The scaling behaviors predicted by the above out-of-equilibrium FSS
theory are strongly supported by numerical analyses of the quantum
Ising chain along the generalized KZ protocols outlined in
Sec.~\ref{KZprot}. Some results are reported in in Fig.~\ref{dfss},
for the magnetization and the adiabaticity function as a function of
the scaling time $\Theta = t / t_s^{\kappa}$ (where $\kappa=1/2$),
keeping the scaling variables $\Upsilon\equiv t_s/L^\zeta$ and
$\Phi\equiv h L^{y_h}$ (where $\zeta=2$ and $y_h=15/8$) and the
initial value $w_i$ fixed.  The evident collapse of the data with
increasing $L$ nicely confirm the predicted out-of-equilibrium FSS
behaviors, i.e. Eq.~(\ref{magsca2}) for the magnetization and
Eq.~(\ref{wsca2}) for the adiabaticity function. We have also checked
that the asymptotic FSS functions do not depend on the initial $w_i<0$
(keeping it fixed with increasing $L$). Analogous results are obtained
for other values of the scaling variables $\Upsilon$ and $\Phi$, and other
monitoring observables, such as the excitation energy (\ref{etdiff}).

\subsection{Scaling in the thermodynamic limit}
\label{thlim}

The scaling behavior in the infinite-size {\em thermodynamic} limit
can be straightforwardly obtained by taking the $L\to\infty$ limit of
the FSS equations, therefore in the limit $\Upsilon\to 0$ keeping
$\Theta$ and $\Sigma$ fixed, cf. Eq.~(\ref{scalvars}). Equivalently,
one may set $b=\lambda$ in the homogeneous scaling laws
(\ref{Oscadynwt}) and (\ref{g12scadynwt}), and then send
$L/\lambda\to\infty$.  Thus, taking the large-$t_s$ limit keeping the
initial value $w_i$ fixed, we expect the asymptotic out-of-equilibrium
scaling behavior
\begin{eqnarray}
&&M(t,t_s,w_i,h,L\to\infty) \approx
  \lambda^{-y_l} {\cal M}_\infty(\Theta,\Sigma)\,,
  \label{magsca3}\\
  &&   G({\bm x},t,t_s,w_i,h,L\to\infty) \approx
  \lambda^{-2y_l}\, {\cal
      G}_\infty({\bm x}/\lambda,\Theta,\Sigma)\,,\qquad
  \label{G2sca3}
\end{eqnarray}
where $\Sigma = h \, \lambda^{y_h}$ was defined in
Eq.~(\ref{scalvars}).  One can easily derive the relation
\begin{eqnarray}
  &&{\cal M}_\infty(\Theta,\Sigma) =
  \lim_{\Upsilon\to 0} \Upsilon^{y_l/\zeta} 
{\cal M}(\Upsilon, \Theta,\Upsilon^{-y_h/\zeta} \Sigma)\,,
\label{minfrel}
\end{eqnarray}
where ${\cal M}$ is the scaling function entering Eq.~(\ref{magsca2}).
An analogous relation can be also written for the two-point function
$G$.

Concerning the excitation-energy density $D_e$, cf. Eq.~(\ref{dedef}),
we expect that in the infinite-volume limit
\begin{eqnarray}
  D_e(t,t_s,w_i,h,L\to\infty)
  \approx \lambda^{-(z+d)} \, {\cal E}_\infty(\Theta,\Sigma)\,.
\label{esscaw3}
\end{eqnarray}
This is analogous to the scaling behavior of the excitation-energy
density reported in Ref.~\onlinecite{RDZ-19}.

\subsection{Scaling corrections}
\label{scacorr}

The out-of-equilibrium FSS limit is expected to be approached with
power-law suppressed corrections.  Scaling corrections to the
asymptotic scaling behaviors arises for finite time scales $t_s$ and
finite size $L$, in particular when they are moderately large.

The sources of scaling corrections when approaching the
out-of-equilibrium FSS are expected to include those that are already
present at equilibrium.  The irrelevant RG perturbations at the fixed
point associated with the $(d+1)$-dimensional Ising fixed point are
sources of scaling corrections.  The contributions of the leading
irrelevant RG perturbation are generally suppressed by a power law, as
$\xi^{-\omega}$ (where $\xi$ is the diverging correlation length, or the
KZ length scale $\lambda$).~\cite{PV-02,RV-21} These corrections are
expected to be the leading ones in two-dimensional quantum Ising
systems, for which $\omega\approx 0.83$, see
e.g. Ref.~\onlinecite{RV-21} and references therein.  For
one-dimensional quantum Ising systems where
$\omega=2$~\cite{CHPV-02,CCCPV-00,CH-00,CPRV-98,CPV-14}, other
contributions may become more relevant, such as those arising from
analytical backgrounds to the critical
behavior~\cite{PV-02,RV-21}. Earlier studies of out-of-equilibrium
behaviors of the Ising chains along standard KZ protocols~\cite{RV-21}
found that the leading corrections are typically $O(1/L)$, or
equivalently $O(1/\lambda)$, for the observables considered in this
paper.  We therefore expect that the asymptotic out-of-equilibrium FSS
of quantum Ising chains along the generalized KZ protocol is
approached with $O(1/L)$ corrections.

The numerical results for the Ising chains are in substantial
agreement with the above analysis of the expected scaling corrections.
The approach to the large-$t_s$ (or equivalently large-$L$) asymptotic
behavior turns out to be substantially consistent with an $O(1/L)$
suppression of the corrections, see e.g. the results reported in
the insets of Fig.~\ref{dfss}.

\section{Approach to the adiabatic regime}
\label{adiabatic}

In this section we focus on the adiabatic limits that can be obtained
within the out-of-equilibrium FSS scenario outlined in the previous
section.  Within generalized KZ protocols, there are essentially two
roads to adiabaticity: one is related to the finite size $L$ and the
other one to the symmetry-breaking perturbation.

In the limit $\Upsilon=t_s/L^{\zeta}\to\infty$, the evolution as a function of
$w(t)=t/t_s$ becomes adiabatic, i.e. it passes through the ground
states associated with the instantaneous values $w(t)$ (when starting
from the ground state for the initial value $w_i$).  Indeed, since the
finite size $L$ guarantees the presence of a gap between the lowest
states, even at the critical point, the critical point can be
adiabatically crossed if $\Upsilon\to\infty$, passing through the
ground states of the finite-size system for $w(t)$. The adiabatic
evolution across the transition point is prevented only when
$L\to\infty$ (before the limit $t_s\to\infty$) and $h=0$.

Another adiabatic limit is related to the presence of the longitudinal
field $h$, due to the fact that, in the presence of a nonvanishing
longitudinal field $h$, the gap never closes, indeed at the critical
point $w=w_c$ it remains finite in the large-$L$ limit, behaving as
$\Delta \sim |h|^{z/y_h}$ for small $|h|$.

\subsection{The adiabatic limit}
\label{upsinfty}

Within the FSS framework, the adiabatic limit is achieved by taking
the $\Upsilon\to\infty$ limit keeping $K$ fixed,
cf. Eq.~(\ref{scalvars}).  Therefore in the adiabatic limit
$\Upsilon\to\infty$ the scaling functions must tend to those of the
equilibrium FSS, i.e. for a generic observable in the
$\Upsilon\to\infty$ limit
\begin{eqnarray}
  O(t,t_s,w_i,h,L) 
  \approx L^{-y_o}
  {\cal
    O}_{\rm eq}(K,\Phi)\,,
  \label{magscaoeq}
\end{eqnarray}
where ${\cal O}_{\rm eq}(K,\Phi)=\widetilde{\cal
  O}(\Upsilon\to\infty,K,\Phi)$ is the equilibrium scaling function
associated with the ground state of the finite-size system across the
transition, entering the equilibrium FSS of the observable $O$,
i.e.~\cite{RV-21,CPV-14}
\begin{equation}
O(w,h,L)
\approx L^{-y_o} {\cal O}_{\rm eq}(w L^{y_w},h L^{y_h})\,.
\label{oeqsca}
\end{equation}

\begin{figure}[!htb]
  \includegraphics[width=0.95\columnwidth]{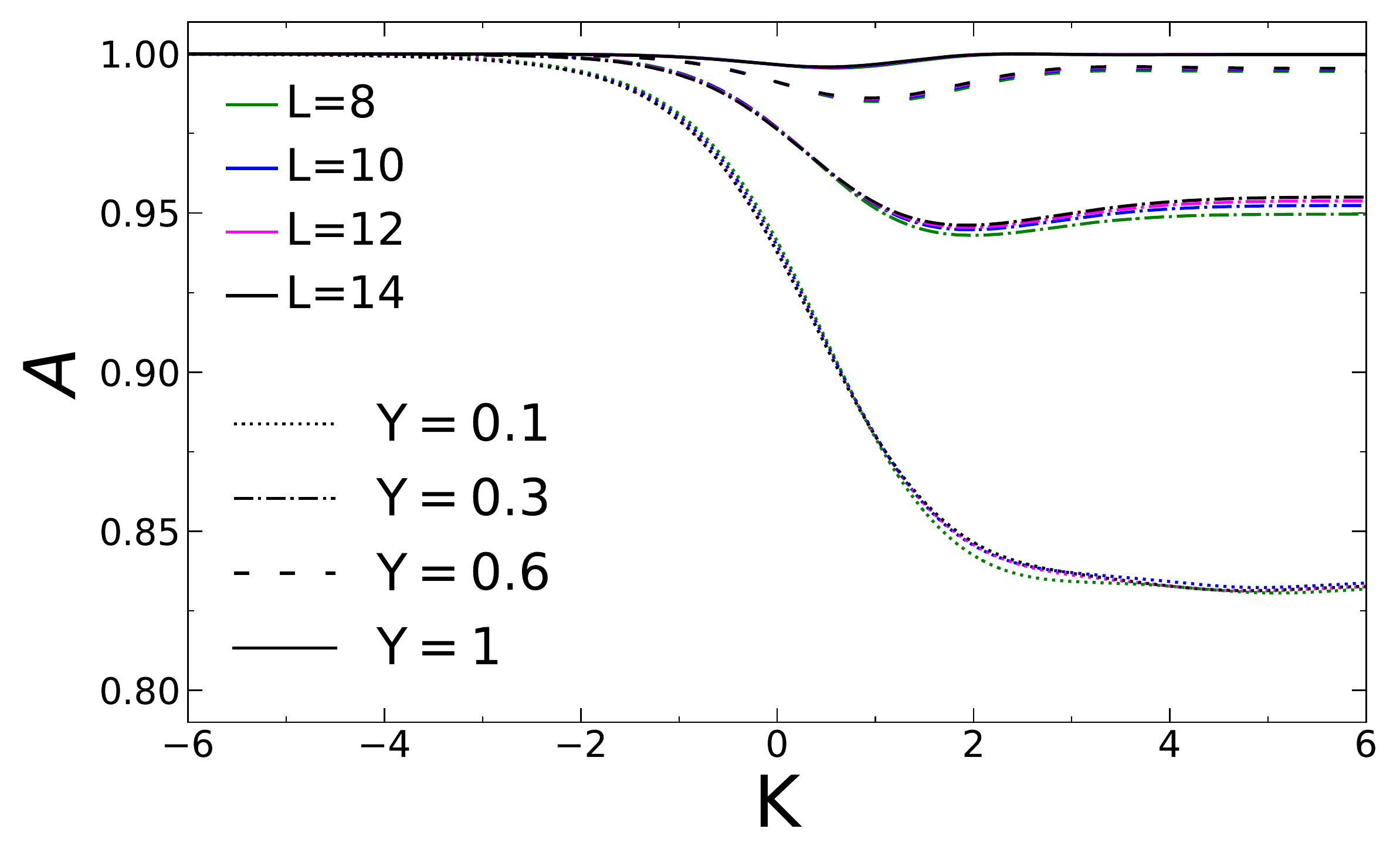}
  \caption{The adiabaticity function $A$ of the quantum Ising chain
    along generalized KZ protocols, versus $K=w(t) L^{y_w}$ at fixed
    $\Phi=h L^{y_h}=1$ and $w_i=-1$, for various values of $\Upsilon$
    and $L$. The curves for different values of $\Upsilon$ appear to
    follow different asymptotic curves, which show the expected
    approach to the adiabatic value $A=1$ with increasing $\Upsilon$,
    in agreement with Eq.~(\ref{adlim}). Note also the nonmonotonic
    dependence on $K$ for sufficiently large $\Upsilon$, likely due to a
    nontrivial interplay between effects related to the finite size and
    the presence of an external symmetry-breaking longitudinal field.}
         \label{adialim}
\end{figure}

The adiabaticity function must behave trivially in the adiabatic
limit, so that the large-$\Upsilon$ limit of the scaling function
$\widetilde{A}$ entering Eq.~(\ref{wsca2}) must be
\begin{equation}
\lim_{\Upsilon\to \infty} \widetilde{\cal A}(\Upsilon,K,\Phi)=1\,.
\label{adlim}
\end{equation}
This limit is also supported by the numerical analyses within the
quantum Ising chain, see e.g.  Fig.~\ref{adialim} where the scaling
curves appear to approach the adiabatic limit (\ref{adlim}) with
increasing $\Upsilon$.  Analogously, the excitation energy must vanish
in the adiabatic limit, by construction, i.e.
\begin{equation}
  \lim_{\Upsilon\to \infty} \widetilde{\cal E}(\Upsilon,K,\Phi)
  =0\,.
\label{adlime}
\end{equation}

\subsection{Power-law approach to the adiabatic regime}
\label{appadreg}

The power-law approach to the adiabatic limit of the
out-of-equilibrium FSS behaviors, i.e. for large values of $\Upsilon$,
can be inferred by exploiting the adiabatic perturbation theory, see
e.g. Refs.~\onlinecite{Schiff-book,ROP-08,DP-10,PSSV-11}.

By expanding the state $|\Psi(t)\rangle$ in terms of the instantaneous
Hamiltonian eigenstates $|\Psi_n[w(t)]\rangle$ (assuming no degeneracy
and that $n=0$ is the lowest eigenstate), 
\begin{equation}
  |\Psi(t)\rangle = \sum_{n\ge 0} a_n(t) |\Psi_n[w(t)]\rangle\,,
  \label{adiaexp}
\end{equation}  
we can write the overlap with the ground state associated with the
instantaneous value of $w(t)$ as
\begin{eqnarray}
  |\langle \, \Psi_0[w(t)] \, | \, \Psi(t) \,\rangle|^2 =
  1 - \sum_{n>0} |a_n(t)|^2 \,.
  \label{superexp}
\end{eqnarray}
We can use the first-order adiabatic approximation to estimate the
amplitudes $a_n(t)$.  This is essentially justified by the fact that
the out-of-equilibrium FSS limit implies $t_s\gg L^{-z}$,
cf. Eq.~(\ref{tsde}).  Using the general results, see e.g.
Refs.~\onlinecite{DP-10,DGP-10,ROP-08}, at the time $t=0$
corresponding to the critical point when $w=w_c=0$, we obtain
\begin{eqnarray}
|\langle \, \Psi_0[w(0)] \, | \, \Psi(0) \,
  \rangle|^2 -1
      \approx  t_s^{-2} \sum_{n>0}{
    |\langle n| \hat{H}_w |0\rangle|^2 \over (\Delta E_n)^4}\,,
\label{tsbeha} 
\end{eqnarray}
where $\Delta E_n \equiv E_n(w=0) - E_0(w=0)$ is the energy difference
between the $n^{\rm th}$ level and the ground state at $w=0$.  Then,
using the expected scaling behaviors at CQTs,~\cite{RV-21}
\begin{eqnarray}
\Delta E_n \sim L^{-z}\,,\qquad
\langle
  n| \hat{H}_w |0\rangle \sim L^{d-y_t}\,,
  \label{htsca}
\end{eqnarray}
where $y_t=d+z-y_w$, we obtain 
\begin{eqnarray}
  A^2 - 1 \sim t_s^{-2} L^{2(z+y_w)} = \Upsilon^{-2}\,.
  \label{adiascaa}
\end{eqnarray}
This result is expected to extend to generic values of $K$ along the
out-of-equilibrium evolution (we also checked it numerically, see
below).  Therefore, the out-of-equilibrium FSS keeping $K$ and $\Phi$
fixed, cf. Eq.~(\ref{wsca2}), should asymptotically behave as
\begin{eqnarray}
\widetilde{\cal A}(\Upsilon\to\infty,K,\Phi) \approx \, 
 1 -  a \, \Upsilon^{-2}\,,
\label{Asca4}
\end{eqnarray}
to match the asymptotic Eq.~(\ref{tsbeha}), where the coefficient $a$
generally depends on $K$ and $\Phi$.  Analogously for the excitation
energy (\ref{etdiff}), using the first-order adiabatic approximation
one arrives at the asymptotic power-law suppression
\begin{eqnarray}
\widetilde{\cal E}(\Upsilon\to\infty,K,\Phi)
    \sim \Upsilon^{-2}\,.\qquad
    \label{wsca4}
  \end{eqnarray}
We remark that the $O(t_s^{-2})$ behavior in the adiabatic limit is
obtained assuming a discrete spectrum, which is appropriate in
finite-size systems, and expected to extend to the out-of-equilibrium
FSS limit, in particular when the out-of-equilibrium FSS behavior
approaches the equilibrium FSS, in the limit $\Upsilon\to\infty$
keeping $K$ and $\Phi$ constant.

\begin{figure}[!htb]
      \includegraphics[width=0.95\columnwidth]{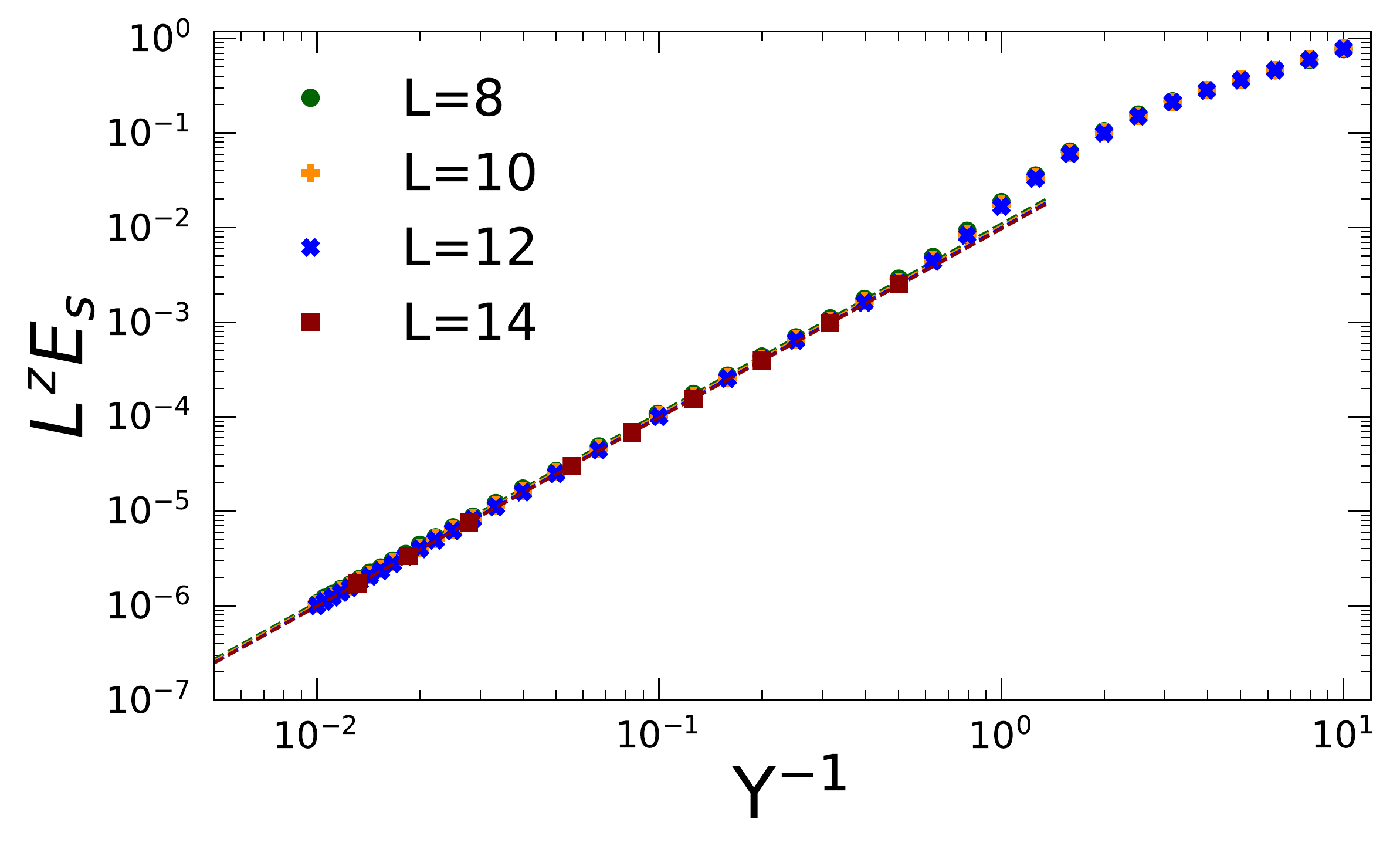}
  \includegraphics[width=0.95\columnwidth]{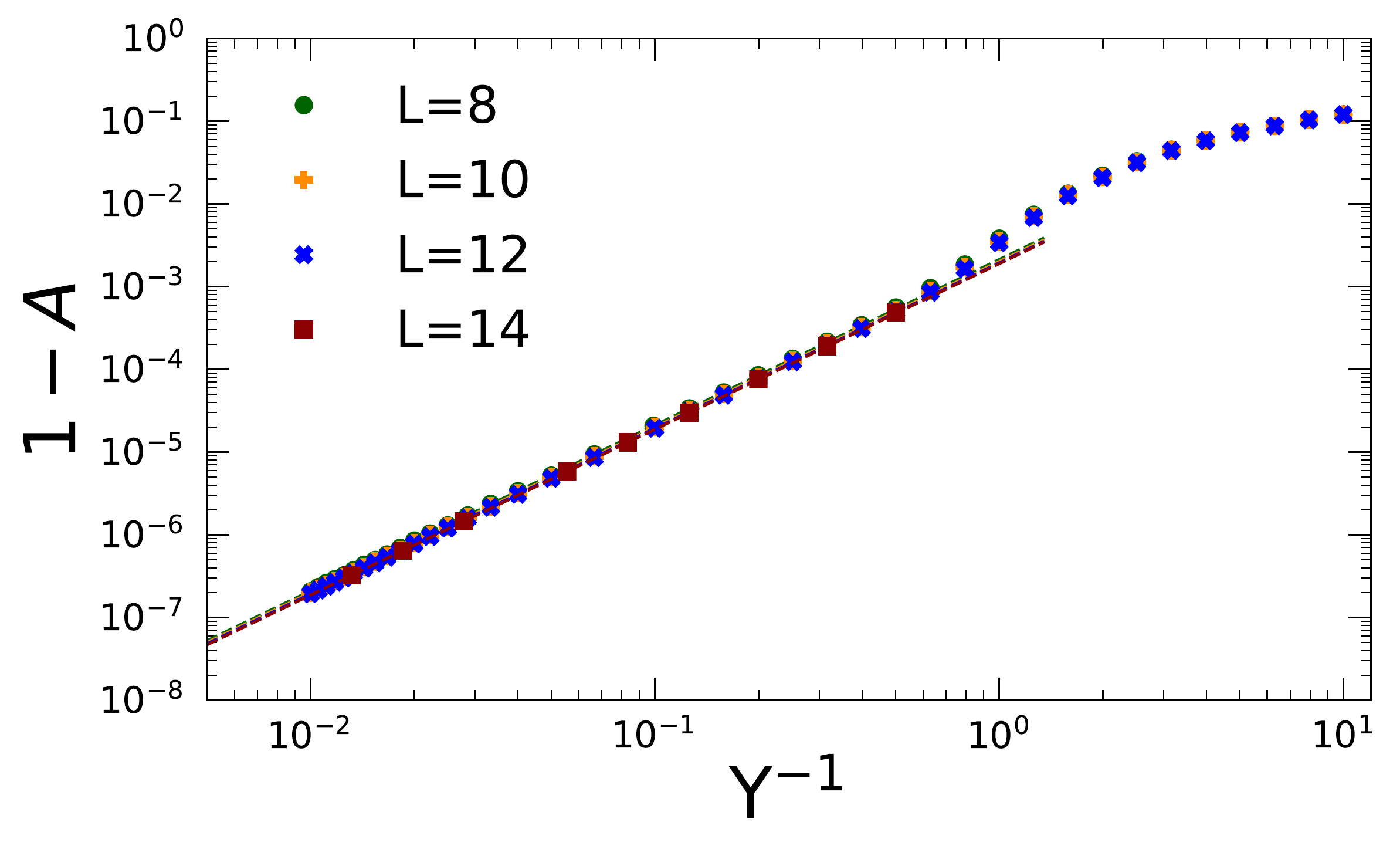}
  \caption{Log-log plots of the the adiabaticity function (bottom) and
    the excitation energy (top) for fixed scaling variables $K\equiv
    w(t)L^{y_w}=1$ and $\Phi= h L^{y_h}=1$, and fixed $w_i=-1$, versus
    $\Upsilon^{-1}=L^\zeta/t_s$.  The data for different values of $L$
    are hardly distinguishable, demonstrating a good convergence to
    the corresponding asymptotic behavior.  The approach to the
    adiabatic regime for large values of $\Upsilon$ turns out to be
    perfectly consistent with the Eqs.~(\ref{Asca4}) and
    (\ref{wsca4}), predicting an asymptotic $O(\Upsilon^{-2})$
    suppression of the residual nonadiabatic effects (in both figures
    the dashed line shows a linear fit to $b \Upsilon^{-2}$ of the
    data for the largest available values of $\Upsilon$).  Note that
    the asymptotic $\Upsilon^{-2}$ decay is observed for
    $\Upsilon\gtrsim 1$.  }
     \label{adiaappro}
\end{figure}

The above power-law approach to adiabaticity are confirmed by
numerical analyses on the quantum Ising chain. Some results for the
adiabaticity function and the excitation energy along the KZ protocol
are shown in Fig.~\ref{adiaappro}. Analogous results are obtained for
other values of the scaling variables.  Their behaviors clearly
confirm the $\Upsilon^{-2}$ power-law suppression of nonadiabaticity
in the adiabatic limit $\Upsilon\to\infty$, cf. Eqs.~(\ref{Asca4}) and
(\ref{wsca4}).

\subsection{The adiabatic limit due to the longitudinal field}
\label{adiah}

In the presence of an external longitudinal field, adiabaticity is
generally recovered in the limit $t_s\to\infty$ when keeping the
longitudinal field $h\neq 0$ fixed.  Indeed, in the presence of a
longitudinal field $h$, the gap does not close at the critical point
$w=w_c$, i.e. $\Delta(w=w_c,h)\sim |h|^{z/y_h}$, and therefore the KZ
protocol can follow an adiabatic evolution in the limit of large time
scale $t_s\to \infty$, even in the thermodynamic limit.  Therefore,
the out-of-equilibrium FSS outlined in Sec.~\ref{qfssoneway} must also
have a corresponding adiabatic limit, which must be realized when
$|\Phi|\equiv |h| L^{y_h}\to\infty$.  Without loss of generality, we
assume $h>0$, thus $\Phi>0$ and $\Sigma>0$, in the following.
 
To study the $\Phi\to\infty$ limit, it is useful to introduce
the length scale associated with the longitudinal field $h$, i.e.
\begin{equation}
  \lambda_h = h^{-1/y_h} = \Phi^{-1/y_h} L\,.
  \label{lambdah}
\end{equation}
Since $\Phi$ can be written as $\Phi=(L/\lambda_h)^{y_h}$, the
large-$\Phi$ limit corresponds to $L\gg \lambda_h$.  Moreover, we
introduce the scaling variable
  \begin{equation}
    \widetilde\Upsilon = t_s\, \lambda_h^{-\zeta} = \Sigma^{\zeta/y_h} =
    \Upsilon
  \Phi^{\zeta/y_h}\,.
    \label{wupsilon}
  \end{equation}
Note that the adiabatic limit $\Upsilon\to\infty$ is equivalent to
$\widetilde{\Upsilon} \to \infty$ when we keep $\Phi>0$ constant.
However, in the limit $\Phi\to\infty$, $\widetilde{\Upsilon}\to\infty$
even though $\Upsilon$ is kept finite.

  We expect that the scaling functions for adiabaticity function and
  excitation energy become trivial for $\Phi\to\infty$, i.e.
  \begin{eqnarray}
    \lim_{\Phi\to \infty} \widetilde{\cal A}(\Upsilon,K,\Phi) = 1\,,
    \quad \lim_{\Phi\to\infty} \widetilde{\cal
    E}(\Upsilon,K,\Phi)=0\,.
    \label{wsca3}
  \end{eqnarray}
  The adiabatic limit of the
  magnetization is less trivial, indeed we expect that in the limit
  $\Phi\to\infty$
\begin{eqnarray}
  \widetilde{\cal M}(\Upsilon,K,\Phi\to\infty) =
            {\cal M}_{\rm eq}(K,\Phi\to\infty)\,,
  \label{asyphibeh}
\end{eqnarray}
where ${\cal M}_{\rm eq}(K,\Phi)$ is the scaling function entering the
equilibrium FSS~\cite{CPV-14,RV-21}, defined as in Eq.~(\ref{oeqsca})
with $y_o=y_l$.  By matching its asymptotic behavior for
$\Phi\to\infty$ with the power-law $M\sim h^{1/\delta}$ at the
critical point $w=w_c$ and for finite $h$, one can easily derive the
asymptotic behavior ${\cal M}_{\rm eq}(K,\Phi\to\infty)\sim
\Phi^{1/\delta}$.  Note that for the Ising chain $\delta=y_h/y_l =
15$, thus the $O(\Phi^{1/\delta})$ divergence turns out to be very
slow.

\begin{figure}[!htb]  
  \includegraphics[width=0.95\columnwidth]{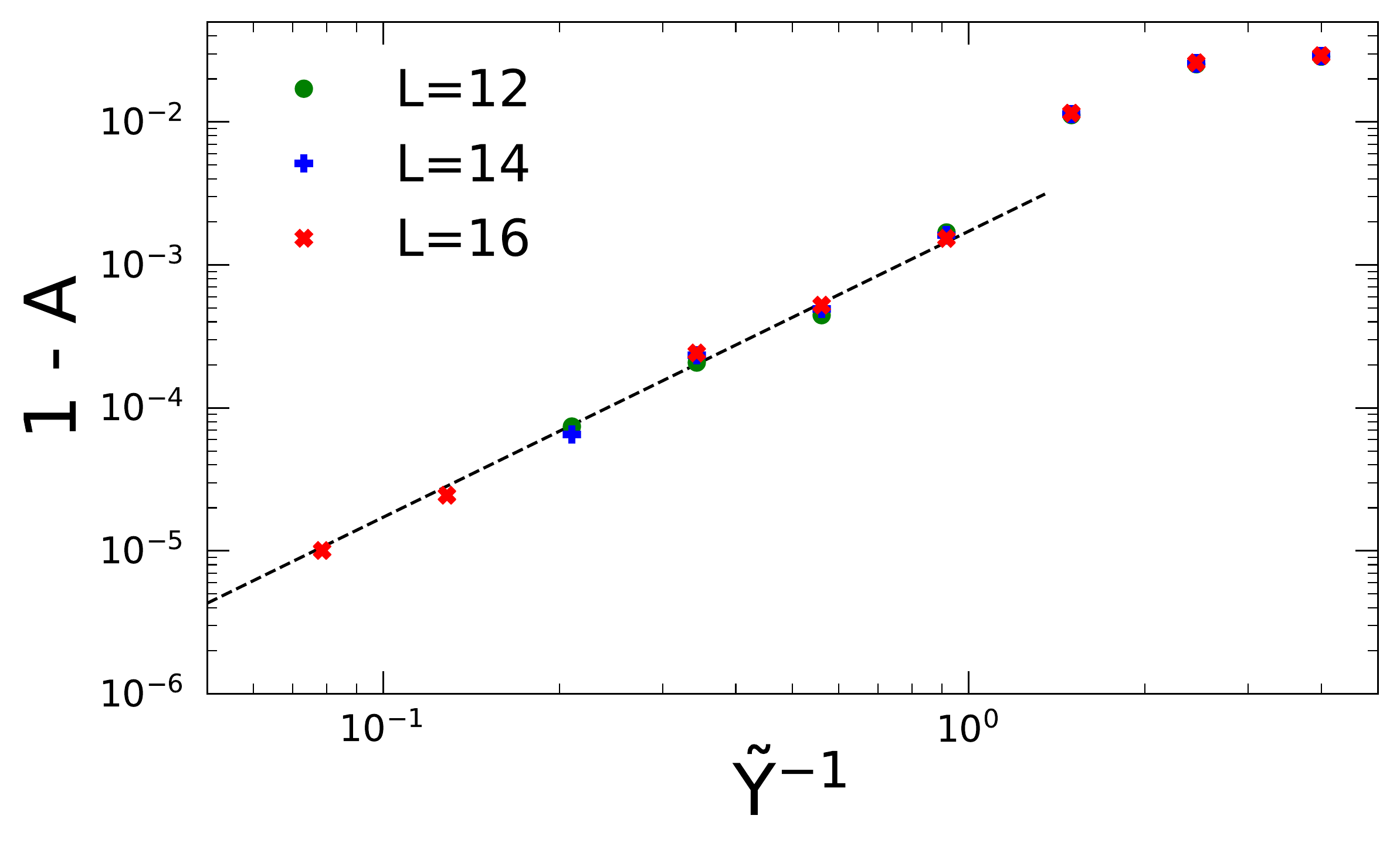}
  \includegraphics[width=0.95\columnwidth]{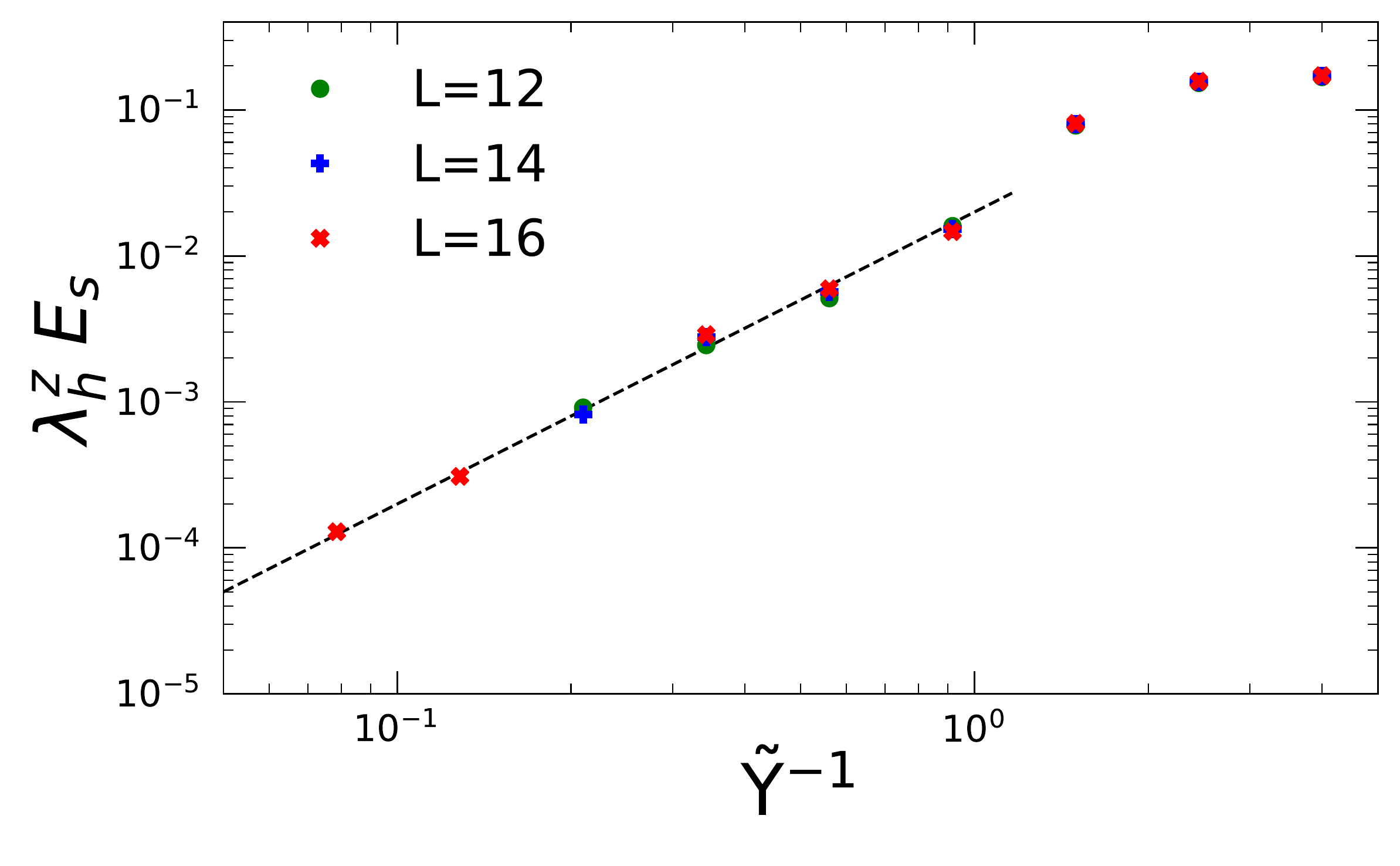}
  \caption{ Log-log plots of the the adiabaticity function $A$ (more
    precisely of $1-A$, top figure) and the energy excitation $E_s$
    (actually the product $\lambda_h^z E_s$, bottom figure) for fixed
    scaling variables $K\equiv w(t)L^{y_w}=0$, $\Upsilon=0.25$, fixed
    $w_i=-1$, versus $\widetilde{\Upsilon}^{-1}$. Data for different values
    of $L$ show a good convergence with increasing $L$.  The resulting
    scaling behaviors in the large $\widetilde\Upsilon$ limit agree
    with Eqs.~(\ref{Asca5}) and (\ref{wsca5}), predicting an
    asymptotic $O(\widetilde\Upsilon^{-2})$ suppression of the
    residual nonadiabatic contributions (the dashed lines show the
    $\widetilde\Upsilon^{-2}$ behavior in the log-log plots).  Such a
    power-law approach is approximately obseved for
    $\widetilde\Upsilon \gtrsim 1$, with some oscillations of
    relatively small amplitude (whose origin is not clear).  }
     \label{adiaphitoinfty}
\end{figure}

Let us now discuss the approach to the adiabatic limit arising from
the presence of a nonzero $h$, i.e. for $\Phi\to\infty$ or
equivalently $\widetilde{\Upsilon}\to\infty$ when keeping $\Upsilon$
fixed. We recall that, in the presence of a nonzero $h$, the gap is
always nonzero, its minimum at $w=w_c$ being $\Delta\sim h^{z/y_h}$.
To discuss the adiabatic limit arising from a nonzero $h$, it is
convenient to focus on the adiabaticity function $A$ and the
excitation energy $E_s$ or the corresponding density $D_e$, in that
they have a trivial adiabatic limit by construction. Using analogous
arguments to those outlined for the adiabatic limit
$\Upsilon\to\infty$ in Sec.~\ref{appadreg}, and in particular the fact
that the adiabaticity violations are generally $O(t_s^{-2})$ within
the first-order adiabatic approximation, the approach to the adiabatic
limit is expected to be characterized by the power law
\begin{eqnarray}
  A(t,t_s,w_i,h,L) \approx \widetilde{\cal
    A}(\Upsilon,K,\Phi\to\infty) \approx 1 - a
  \widetilde{\Upsilon}^{-2}\,,
    \label{Asca5}
\end{eqnarray}
where $\widetilde{\Upsilon} = \Upsilon \Phi^{\zeta/y_h}\to\infty$, and
the factor $a$ generally depends on the other scaling variables $K$
and $\Upsilon$.

Numerical results for the quantum Ising chain confirm
the above asymptotic behavior in large-$\Phi$ limit, as for example
shown by the data reported in the top Fig.~\ref{adiaphitoinfty},
obtained keeping the scaling variables $K$ and $\Upsilon$ fixed (in
particular for $K\equiv w(t)L^{y_w}=0$ corresponding to $t=0$, and
$\Upsilon=0.25$, and also fixed $w_i=-1$, analogous results have been
obtained for other values of $\Upsilon$ and $K$).  Again the data for
different values of $L$ show a good convergence to the corresponding
asymptotic large-$L$ scaling behavior, which agrees with the predicted
$\widetilde\Upsilon^{-2}$ power-law approach to the adiabatic limit
$A=1$.

To derive the asymptotic behavior of the energy excitation, and
consistently perform the large-$\Phi$ limit of the scaling
Eq.~(\ref{essca}), we must replace the $L^{-z}$ prefactor in the
r.h.s. of Eq.~(\ref{essca}) with a corresponding expression in terms
of $h$ and $\Phi$, such as $L^{-z} = \lambda_h^{-z} \Phi^{-z/y_h}$,
where $\lambda_h$ is the length scale associated with the longitudinal
field $h$, cf. Eq.~(\ref{lambdah}).  Then we expect
\begin{eqnarray}
  \lambda_h^{z} E_s(t,t_s,w_i,h,L) \approx 
  \Phi^{-z/y_h} \widetilde{\cal E}(\Upsilon,K,\Phi) \approx c
  \widetilde{\Upsilon}^{-2}\,,\quad
    \label{wsca5}
\end{eqnarray}
where $c$ generally depends on the scaling variables $K$ and
$\Upsilon$.  Some numerical results for the quantum Ising chain are
reported in the bottom Fig.~\ref{adiaphitoinfty}, where the 
power-law approach (\ref{wsca5}) to the adiabatic limit of the
excitation energy is clearly observed.

We finally note that the $\Phi\to\infty$ limit can be also seen as an
inifinite volume limit, in that it corresponds to
$L/\lambda_h\to\infty$.  Let us consider the energy-excitation density
(\ref{dedef}), which is expected to have a well defined
infinite-volume limit.  By simple manipulations of Eq.~(\ref{essca}),
one can write its scaling behavior as
\begin{eqnarray}
  D_e(t,t_s,w_i,h,L) \approx \lambda_h^{-(z+d)} \, \widehat{\cal
    E}(\widetilde{K},\Sigma,L/\lambda_h)\,,
\label{esscaw3}
\end{eqnarray}
where $\widetilde{K}=w(t)\lambda_h^{y_w}$, $\Sigma = h
\lambda^{y_h}=(\lambda/\lambda_h)^{y_h}=\widetilde{\Upsilon}^{y_h/\zeta}$,
and $\lambda$ and $\lambda_h$ are respectively the length scales
defined in Eq.~(\ref{lambdats}) and Eq.~(\ref{lambdah}).  Then we take
the limit $L/\lambda_h\to\infty$ keeping $\widetilde{K}$ and $\Sigma$
fixed, thus, assuming that such a limit is regular,
\begin{eqnarray}
  D_e(t,t_s,w_i,h,L\to\infty)
  \approx \lambda_h^{-(z+d)} \,
  \widehat{\cal E}_\infty(\widetilde{K},\Sigma)\,.
\label{esscaw3b}
\end{eqnarray}
Then, we note that taking the further limit $\Sigma,
\widetilde{\Upsilon}\to\infty$ keeping $\widetilde{K}$ fixed
corresponds to an adiabatic limit, so that
\begin{eqnarray}
  \lim_{\Sigma\to\infty} \widehat{\cal E}_\infty(\widetilde{K},\Sigma) \to 0\,.
\label{esscaw6}
\end{eqnarray}
Assuming that the asymptotic behavior (\ref{wsca5}) persists in the
infinite-volume limit corresponding to $\Upsilon\to 0$ (notice that
this is possible because $\Upsilon$ ad $\widetilde\Upsilon$ are not
anymore related when $\Phi\to\infty$), we also expect
\begin{eqnarray}
\widehat{\cal E}_\infty(\widetilde{K},\Sigma\to\infty) \sim
  \widetilde{\Upsilon}^{-2}\,,
\label{esscaw7}
\end{eqnarray}    
where $\widetilde{\Upsilon} = \Sigma^{\zeta/y_h} = t_s h^{\zeta/y_h}$.
One can easily check that this asymptotic behavior agrees with the
results reported in Ref.~\onlinecite{RDZ-19} for $t=0$, corresponding to
$\widetilde{K}=0$. Indeed, we obtain
\begin{equation}
  D_e \sim \lambda_h^{-(z+d)} \widetilde\Upsilon^{-2} =
  t_s^{-2} h^{(d-z-2y_w)/y_h}\,,
  \label{derdz}
  \end{equation}
corresponding to $D_e \sim t_s^{-2} h^{-16/15}$ for the Ising chain.

\section{Extended KZ protocols}
\label{bothtimedep}

We now consider a KZ protocol in which both Hamiltonian parameters $w$
and $h$ are assumed to be time dependent and cross their critical
values with a linear time dependence.  More precisely, we extend the
time dependence of the Hamiltonian (\ref{hlamt}) to
\begin{eqnarray}
&&\hat H[w(t),h(t)] = \hat H_{c} + w(t) \, \hat H_{w} + h(t) \hat H_h\,,
\label{hlamt2}\\
&& w(t) = t/t_s\,,\qquad h(t) = t/t_{s,h}\,.
\nonumber
\end{eqnarray}
In this case we allow for different time scales in the time
dependence of $w$ and $h$, but both of them take their critical value
at $t=0$.  We consider a KZ protocol analogous to that outlined in
Sec.~\ref{KZprot}, but allowing for the time changes of $h$ too,
starting at $t=t_i<0$ from the ground state associated with $w(t_i)$
and $h(t_i)$, and then unitarily evolving using the Hamiltonian $\hat
H[w(t),h(t)]$.

We first consider the simplest case in which $t_{s,h} = c \,t_s$ where
$c$ is a finite constant.  To address the resulting out-of-equilibrium
scaling behavior, one may again exploit homogenous scaling laws
analogous to those reported in Eq.~(\ref{Oscadynwt}) and
(\ref{g12scadynwt}), with one scaling variable for each time-dependent
parameters $w(t)$ and $h(t)$, i.e. $K = w(t) L^{y_w}$ and $K_h = h(t)
L^{y_h}$.  However, since the RG dimensions of the even and odd model
parameters $w$ and $h$ differ, i.e. $y_w \neq y_h$, and their time
dependence is controlled by a unique time scale $t_s\sim t_{s,h}$, we
cannot consistently rescale them to get nontrivial FSS behaviors
keeping $K$ and $K_h$ fixed to any finite nonzero value. In
particular, if we keep $K=(t/t_s)L^{y_w}$ fixed in the simultaneous
large $t_s$ and $L$ limits, then $K_h\to \infty$ due to the fact that
$y_h>y_w$ generally. Thus no asymptotic FSS is expected to emerge when
keeping $K$ fixed. On the other hand, if we keep $K_h = (t/t_{s,h})
L^{y_h}$ fixed in the FSS limit, then $K\to 0$.  Therefore, the
emerging scaling behavior is analogous to that obtained in KZ
protocols keeping $w=w_c=0$ fixed and varying only the
symmetry-breaking parameter $h(t)$, see e.g.  Ref.~\onlinecite{TV-22}.

In conclusion, we do not expect to observe further interesting
out-of-equilibrium scaling phenomena when we allow for variations of
both the even and odd Hamiltonian parameters $w$ and $h$.  If we
insist on varying both parameters, to obtain a nontrivial scaling
behavior we need to assume different time scales, tuned so that
$t_{s,h}/t_{s} \sim L^{y_h-y_w}$, to have $K \sim K_h$.

\section{Conclusions}
\label{conclu}

We have studied the effects of static homogenous symmetry-breaking
perturbations in the out-of-equilibrium quantum dynamics of many-body
systems driven across a CQT by a time-dependent symmetry-preserving
parameter. Out-of-equilibrium FSS behaviors emerge along generalized
KZ protocols even in the presence of the symmetry-breaking
perturbation, described in Sec.~\ref{KZprot}.  For this purpose we
exploit RG scaling frameworks implemented in terms of asymptotic
homogenous scaling laws, cf. Eqs.~(\ref{Oscadynwt}) and
(\ref{g12scadynwt}), that are expected to hold along the generalized
KZ protocol.  This allows us to develop a generalized
out-of-equilibrium FSS theory, allowing for the symmetry-breaking
perturbation. The emerging out-of-equilibrium FSS scenario in the
presence of a static symmetry-breaking perturbation requires an
appropriate tuning of the corresponding Hamiltonian parameter,
controlled by its RG dimension at the fixed point associated with the
CQT.

As paradigmatic models, we consider the quantum Ising models, whose
CQTs are characterized by the spontaneous breaking of their ${\mathbb
  Z}_2$ symmetry, and two relevant parameters: the symmetry-preserving
transverse field $g$, or $w=g_c-g$, and the symmetry-breaking
longitudinal field $h$, cf. Eq.~(\ref{qisingmodel}).  We consider
generalized KZ protocols, where the out-of-equilibrium dynamics is
induced by variations of the symmetry-preserving parameter $w(t)$
across its critical point, with a linear time dependence $w(t)=t/t_s$,
from the disordered to ordered phase, and a large time scale
$t_s$. Unlike most earlier studies based on KZ protocols, we allow for
a nonzero longitudinal field $h$. We extend the study of
Ref.~\onlinecite{RDZ-19} where the out-of-equilibrium scaling behaviors of
KZ protocols in the presence of a symmetry-breaking perturbation were
discussed in the infinite-volume (thermodynamic) limit.

Within the out-of-equilibrium FSS framework of generalized KZ
protocols in the presence of the symmetry-breaking perturbation, the
adiabatic limit may arise for essentially two reasons, i.e. because of
the finite size (thus the critical gap does not vanish, getting
suppressed as $\Delta\sim L^{-z}$), and because of the longitudinal
field $h$ (which does not allow the gap to vanish even in the
thermodynamic limit, behaving as $\Delta \sim |h|^{z/y_h}$ for
sufficiently small values of $h$). We argue that the nonadiabaticity
of the out-of-equilibrium dynamics gets suppressed by power laws,
which can be inferred from the first-order adiabatic approximation.

The scaling laws obtained within the out-of-equilibrium FSS framework
are confirmed by numerical analyses of generalized KZ protocols in the
quantum Ising chain. We also remark that the out-of-equilibriumn FSS
scenario that we develop can be straightforwardly extended to generic
$d$-dimensional systems undergoing CQTs.

We have also briefly discussed the case in which the generalized KZ
protocol is characterized by the time dependence of both even and odd
parameters $w$ and $h$. Within analogous scaling frameworks, we argue
that no further interesting scaling behaviors emerge when both
parameters are changed with the same time scales. Nontrivial FSS
behaviors may be only obtained by appropriately rescaling the time
scales of $w(t)$ and $h(t)$, proportionally to different powers of the
size.

As already mentioned in the introduction, the FSS approaches generally
simplify the analyses of the universal features of critical behaviors,
essentially because they do not require the further condition $\xi\ll
L$ of the thermodynamic limit.  In this study we have shown that the
out-of-equilibrium FSS scenarios associated with generalized KZ
protocols can be already observed for relatively small systems, with a
few decades of spin operators.  Scaling behaviors requiring a limited
numbers of degrees of freedom may be more easily accessed by
experiments where the coherent quantum dynamics of only a limited
number of particles or spins can be effectively realized, such as
experiments with quantum simulators in laboratories, e.g., by means of
trapped ions~\cite{Islam-etal-11, Debnath-etal-16}, ultracold
atoms~\cite{Simon-etal-11, Labuhn-etal-16}, or superconducting
qubits~\cite{Salathe-etal-15, Cervera-18}. Moreover, our results may
turn out to be particularly relevant for quantum simulations and
quantum computing, where important experimental advances have been
achieved recently, see
e.g. Refs.~\onlinecite{CZ-12,BDN-12,BR-12,AW-12,HTK-12,GAN-14}.

\acknowledgments

We thank Alessio Franchi and Francesco Tarantelli for useful
discussions.


\begin{thebibliography}{99}

\bibitem{Kibble-80} T. W. B. Kibble, Some implications of a
  cosmological phase transition, Phys. Rep. {\bf 67}, 183 (1980).
  

\bibitem{Binder-87} K. Binder, Theory of first-order phase
  transitions, Rep. Prog. Phys. {\bf 50}, 783 (1987).

\bibitem{Bray-94}
  A. J. Bray, Theory of phase-ordering kinetics,
  Adv. Phys. {\bf 43}, 357 (1994).

\bibitem{Zurek-96} W. H. Zurek, Cosmological experiments in condensed
  matter systems, Phys. Rep. {\bf 276}, 177 (1996).

\bibitem{CG-05} P. Calabrese and A. Gambassi, Ageing properties of
  critical systems, J. Phys. A: Math. Gen. {\bf 38}, R133 (2005).

    

  \bibitem{BDS-06} D. Boyanovsky, H. J. de Vega, and D. J. Schwarz,
    Phase transitions in the early and the present universe,
      Annu. Rev. Nucl. Part. Sci. {\bf 56}, 441 (2006).


\bibitem{Dziarmaga-10}
  J. Dziarmaga, Dynamics of a quantum phase transition and relaxation
  to a steady state, Adv. Phys. {\bf 59}, 1063 (2010).

  

      
\bibitem{PSSV-11} A. Polkovnikov, K. Sengupta, A. Silva, and
  M. Vengalattore, Colloquium: Nonequilibrium dynamics of closed
  interacting quantum systems, Rev. Mod. Phys. {\bf 83}, 863 (2011).

\bibitem{Biroli-16} G. Biroli, Slow relaxations and nonequilibrium
  dynamics in classical and quantum systems, in {\em Strongly
    interacting quantum systems out of equilibrium: Lecture notes of
    the Les Houches Summer School}, Aug. 2012, (Oxford University
  Press, 2016).


  
\bibitem{RV-21} D. Rossini and E. Vicari, Coherent and dissipative
  dynamics at quantum phase transitions, Phys. Rep. {\bf 936}, 1
  (2021).


\bibitem{Kibble-76}
  T. W. B. Kibble, 
  Topology of Cosmic Strings and Domains, J. Phys. A {\bf 9}, 1387 (1976).
  

\bibitem{CDTY-91} I. Chuang, R. Durrer, N. Turok, and B. Yurke,
  Cosmology in the Laboratory: Defect Dynamics in Liquid Crystals,
  Science {\bf 251}, 1336 (1991).

\bibitem{BCSS-94} M. J. Bowick, L. Chandar, E. A. Schiff, and
  A. M. Srivastava, The Cosmological Kibble Mechanism in the
  Laboratory: String Formation in Liquid Crystals, Science {\bf 263},
  943 (1994).
  
 \bibitem{BBFGP-96} C. B\"auerle, Yu M. Bunkov, S. N. Fisher,
  H. Godfrin, and G. R. Pickett, Laboratory simulation of cosmic
  string formation in the early Universe using superfluid 3He, Nature
  {\bf 382}, 332 (1996). 
  


\bibitem{WNSBD-08} C. N. Weiler, T. W. Neely, D. R. Scherer,
  A. S. Bradley, M. J. Davis, and B. P. Anderson,
Spontaneous vortices in the formation of Bose–Einstein condensates,
Nature  {\bf 455}, 948 (2008).      


 \bibitem{Ulm-etal-13} S. Ulm, S. J. Ro{\ss}nagel, G. Jacob,
  C. Deg\"unther, S. T. Dawkins, U. G. Poschinger, R. Nigmatullin,
  A. Retzker, M. B. Plenio, F. Schmidt-Kaler, and K. Singer,
  Observation of the Kibble-Zurek scaling law for defect formation in
  ion crystals, Nat. Commun. {\bf 4}, 2290 (2013).

\bibitem{Pyka-etal-13} K. Pyka, J. Keller, H. L. Partner,
  R. Nigmatullin, T. Burgermeister, D. M. Meier, K. Kuhlmann,
  A. Retzker, M. B. Plenio, W. H. Zurek, A. del Campo, and
  T. E. Mehlst\"aubler, Topological defect formation and spontaneous
  symmetry breaking in ion Coulomb crystals, Nat. Commun. {\bf 4},
  2291 (2013).

\bibitem{NGSH-15} N. Navon, A. L. Gaunt, R. P. Smith, and
  Z. Hadzibabic, Critical dynamics of spontaneous symmetry breaking in
  a homogeneous Bose gas, Science {\bf 347}, 167, (2015).


\bibitem{Trenkwalder-etal-16}
A. Trenkwalder, G. Spagnolli,
G. Semeghini, S. Coop, M. Landini, P. Castilho, L. Pezz\`e,
G. Modugno, M. Inguscio, A. Smerzi, and M. Fattori, Quantum phase
transitions with parity-symmetry breaking and hysteresis,
Nat. Phys. {\bf 12}, 826 (2016).
  
\bibitem{Polkovnikov-05} A. Polkovnikov, Universal adiabatic dynamics
  in the vicinity of a quantum critical point, Phys. Rev. B {\bf 72},
  161201(R) (2005).

\bibitem{ZDZ-05}
  W. H. Zurek, U. Dorner, and P. Zoller, Dynamics of a quantum phase
  transition, Phys. Rev. Lett. {\bf 95}, 105701 (2005).


 \bibitem{Dziarmaga-05} J. Dziarmaga, Dynamics of a quantum phase
  transition: Exact solution of the quantum Ising model,
  Phys. Rev. Lett. {\bf 95}, 245701 (2005).


  \bibitem{FFO-07}
  A. Fubini, G. Falci, and A. Osterloh, Robustness of adiabatic
  passage through a quantum phase transition,
  New J. Phys. {\bf 9}, 134 (2007).

  
\bibitem{DGP-10}
  C. De Grandi, V. Gritsev, and A. Polkovnikov,
  Quench dynamics near a quantum critical point,
  Phys. Rev. B {\bf 81}, 012303 (2010);
  Phys. Rev. B {\bf 81}, 224301 (2010).
  
  
\bibitem{GZHF-10}
  S. Gong, F. Zhong, X. Huang, and S. Fan, Finite-time
  scaling via linear driving, New J. Phys. {\bf 12}, 043036 (2010).
  
\bibitem{CEGS-12}
  A. Chandran, A. Erez, S. S. Gubser, and S. L. Sondhi,
  Kibble-Zurek problem: Universality and the scaling limit,
  Phys. Rev. B {\bf 86}, 064304 (2012).

\bibitem{PRV-18-loc} A. Pelissetto, D. Rossini, and E. Vicari,
  Out-of-equilibrium dynamics driven by localized time-dependent
  perturbations at quantum phase transitions, Phys. Rev. B {\bf 97},
  094414 (2018).
  
\bibitem{PRV-18} A. Pelissetto, D. Rossini, and E. Vicari, Dynamic
  finite-size scaling after a quench at quantum transitions,
  Phys. Rev. E {\bf 97}, 052148 (2018).

\bibitem{RV-19-de} D. Rossini and E. Vicari, Scaling of decoherence
  and energy flow in interacting quantum spin systems,
  Phys. Rev. A {\bf 99}, 052113 (2019).  

\bibitem{RDZ-19}
  M. M. Rams, J. Dziarmaga, and W. H. Zurek,
Symmetry Breaking Bias and the Dynamics of a Quantum Phase Transition,
Phys. Rev. Lett. {\bf 123}, 130603 (2019).


\bibitem{Zurek-85} W. H. Zurek, Cosmological Experiments in Superfluid
  Helium?, Nature {\bf 317}, 505 (1985).

\bibitem{PG-08} A. Polkovnikov and V. Gritsev, Breakdown of the
  adiabatic limit in low-dimensional gapless systems, Nature
  Phys. {\bf 4}, 477 (2008).

\bibitem{Dutta-etal-book} A. Dutta, G. Aeppli, B. K. Chakrabarti,
  U. Divakaran, T. F. Rosenbaum, and D. Sen, {\em Quantum phase
    transitions in transverse field spin models: From statistical
    physics to quantum information}, Cambridge University Press
  (2015).
  
\bibitem{PV-16} A. Pelissetto and E. Vicari, Off-equilibrium scaling
  behaviors driven by time-dependent external fields in
  three-dimensional O($N$) vector models, Phys. Rev. E {\bf 93},
  032141 (2016).

\bibitem{PV-17} A. Pelissetto and  E. Vicari, Dynamic off-equilibrium
  transition in systems slowly driven across thermal first-order
  transitions, Phys. Rev. Lett. {\bf 118}, 030602 (2017).
  
 \bibitem{RV-20} D. Rossini and E. Vicari, Dynamic Kibble-Zurek
   scaling framework for open dissipative many-body systems crossing
   quantum transitions, Phys. Rev. Research {\bf 2}, 023611 (2020).

\bibitem{PRV-20} A. Pelissetto, D. Rossini, and E. Vicari, Scaling
  properties of the dynamics at first-order quantum transitions when
  boundary conditions favor one of the two phases, Phys. Rev. E {\bf
    102}, 012143 (2020).

\bibitem{RMAKVE-21}
J. Rysti, J. T. Ma\"akinen, S. Autti,
T. Kamppinen, G. E. Volovik, and V. B. Eltsov, Suppressing the
Kibble-Zurek Mechanism by a Symmetry-Violating Bias,
Phys. Rev. Lett. {\bf 127}, 115702 (2021).
  
      
\bibitem{TV-22} F. Tarantelli and E. Vicari, Out-of-equilibrium
  dynamics arising from slow round-trip variations of Hamiltonian
  parameters across quantum and classical critical points,
  Phys. Rev. B {\bf 105}, 235124 (2022).


   

\bibitem{Damski-05} B. Damski, The simplest quantum model supporting
  the Kibble-Zurek mechanism of topological defect production:
  Landau-Zener transitions from a new perspective,
  Phys. Rev. Lett. {\bf 95}, 035701 (2005).

  
\bibitem{USF-07} M. Uhlmann, R. Sch{\"u}tzhold, and U. R. Fischer,
  Vortex quantum creation and winding number scaling in a quenched
  spinor Bose gas, Phys. Rev. Lett. {\bf 99}, 120407 (2007).

\bibitem{USF-10} M. Uhlmann, R. Sch{\"u}tzhold, and U. R. Fischer,
  System size scaling of topological defect creation in a second-order
  dynamical quantum phase transition, New. J. Phys. {\bf 12}, 095020
  (2010).

\bibitem{NDP-13} T. Nag, A. Dutta, and A. Patra, Quench dynamics and
  quantum information, Int. J. Mod. Phys. B {\bf 27}, 1345036 (2013).
  

 \bibitem{SGCS-97} S. L. Sondhi, S. M. Girvin, J. P. Carini, and
  D. Shahar, Continuous quantum phase transitions,
  Rev. Mod. Phys. {\bf 69}, 315 (1997).

  
\bibitem{Sachdev-book} S. Sachdev, {\em Quantum Phase Transitions},
  (Cambridge University, Cambridge, England, 1999).

 \bibitem{CPV-14} M. Campostrini, A. Pelissetto, and E. Vicari,
  Finite-size scaling at quantum transitions, Phys. Rev. B {\bf 89},
  094516 (2014).


\bibitem{Islam-etal-11}
  R. Islam, E. E. Edwards, K. Kim, S. Korenblit, C. Noh, H. Carmichael,
  G.-D. Lin, L.-M. Duan, C.-C. Joseph Wang, J. K. Freericks, and
  C. Monroe, Onset of a quantum phase transition with a trapped ion
  quantum simulator, Nat. Commun. {\bf 2}, 377 (2011).

\bibitem{Debnath-etal-16} S. Debnath, N. M. Linke, C. Figgatt,
  K. A. Landsman, K. Wright, and C. Monroe, Demonstration of a small
  programmable quantum computer with atomic qubits, Nature {\bf 536},
  63 (2016).

\bibitem{Simon-etal-11} J. Simon, W. S. Bakr, R. Ma, M. E. Tai,
  P. M. Preiss, and M. Greiner, Quantum simulation of
  antiferromagnetic spin chains in an optical lattice, Nature {\bf
    472}, 307 (2011).

\bibitem{Labuhn-etal-16}
  H. Labuhn, D. Barredo, S. Ravets, S. de Leseleuc, T. Macri, T. Lahaye,
  and A. Browaeys, Tunable two-dimensional arrays of single Rydberg atoms
  for realizing quantum Ising models,  Nature {\bf 534}, 667 (2016).  

\bibitem{Salathe-etal-15} Y. Salath{\'e}, M. Mondal, M. Oppliger,
  J. Heinsoo, P. Kurpiers, A. Potocnik, A. Mezzacapo, U. Las Heras,
  L. Lamata, E. Solano, S. Filipp, and A. Wallraff, Digital quantum
  simulation of spin models with circuit quantum electrodynamics,
  Phys. Rev. X {\bf 5}, 021027 (2015).
  
\bibitem{Cervera-18}
  A. Cervera-Lierta, Exact Ising model simulation on a quantum computer,
  Quantum {\bf 2}, 114 (2018).


\bibitem{DRGA-99} S. Ducci, P. L. Ramazza, W. Gonz{\'a}les-Vi{\~ n}as,
  and F. T. Arecchi, Order parameter fragmentation after a
  symmetry-breaking transition, Phys. Rev. Lett. {\bf 83}, 5210
  (1999).
  
\bibitem{MMARK-06} R. Monaco, J. Mygind, M. Aaroe, R. J. Rivers, and
  V. P. Koshelets, Zurek-Kibble mechanism for the spontaneous vortex
  formation in Nb-Al/Al${}_{\rm ox}$/Nb Josephson tunnel junctions:
  New theory and experiment, Phys. Rev. Lett. {\bf 96}, 180604 (2006).

\bibitem{SHLVS-06} L. E. Sadler, J. M. Higbie, S. R. Leslie,
  M. Vengalattore, and D. M. Stamper-Kurn, Spontaneous symmetry
  breaking in a quenched ferromagnetic spinor Bose–Einstein
  condensate, Nature {\bf 443}, 312 (2006).



\bibitem{CWBD-11} D. Chen, M. White, C. Borries, and B. DeMarco,
  Quantum quench of an atomic Mott insulator, Phys. Rev. Lett. {\bf
    106}, 235304 (2011).

\bibitem{Griffin-etal-12} S. M. Griffin, M. Lilienblum, K. T. Delaney,
  Y. Kumagai, M. Fiebig, and N. A. Spaldin, Scaling behavior and
  beyond equilibrium in the hexagonal manganites, Phys. Rev. X {\bf
    2}, 041022 (2012).

  

\bibitem{LDSDF-13} G. Lamporesi, S. Donadello, S. Serafini,
  F. Dalfovo, and G. Ferrari, Spontaneous creation of Kibble-Zurek
  solitons in a Bose-Einstein condensate, Nat. Phys. {\bf 9}, 656
  (2013).

\bibitem{Braun-etal-15} S. Braun, M. Friesdorf, S. S. Hodgman,
  M. Schreiber, J. P. Ronzheimer, A. Riera, M. del Rey, I. Bloch,
  J. Eisert, and U. Schneider, Emergence of coherence and the dynamics
  of quantum phase transitions, Proc. Natl. Acad. Sci. USA {\bf 112},
  3641 (2015).


\bibitem{Chomaz-etal-15} L. Chomaz, L. Corman, T. Bienaim{\'e},
  R. Desbuquois, C. Weitenberg, S. Nascimb{\'e}ne, J. Beugnon, and
  J. Dalibard, Emergence of coherence via transverse condensation in a
  uniform quasi-two-dimensional Bose gas, Nat. Commun. {\bf 6}, 6162
  (2015).

\bibitem{Cui-etal-16} J.-M. Cui, Y.-F. Huang, Z. Wang, D.-Y. Cao,
  J. Wang, W.-M. Lv, L. Luo, A. {del Campo}, Y.-J. Han, C.-F. Li, and
  G.-C. Guo, Experimental trapped-ion quantum simulation of the
  Kibble-Zurek dynamics in momentum space, Sci. Rep. {\bf 6}, 33381
  (2016).

  
\bibitem{Gong-etal-16} M. Gong, X. Wen, G. Sun, D.-W. Zhang, D. Lan,
  Y. Zhou, Y. Fan, Y. Liu, X. Tan, H. Yu, Y. Yu, S.-L. Zhu, S. Han,
  and P. Wu, Simulating the Kibble-Zurek mechanism of the Ising model
  with a superconducting qubit system, Sci. Rep. {\bf 6}, 22667
  (2016).

\bibitem{Anquez-etal-16} M. Anquez, B. A. Robbins, H. M. Bharath,
  M. Boguslawski, T. M. Hoang, and M. S. Chapman,
  Phys. Rev. Lett. {\bf 116}, 155301 (2016).

\bibitem{CFC-16} L. W. Clark, L. Feng, and C. Chin, Universal
  space-time scaling symmetry in the dynamics of bosons across a
  quantum phase transition, Science {\bf 354}, 606 (2016).
  
\bibitem{Keesling-etal-19} A. Keesling, A. Omran, H. Levine,
  H. Bernien, H. Pichler, S. Choi, R. Samajdar, S. Schwartz, P. Silvi,
  S. Sachdev, P. Zoller, M. Endres, M. Greiner, V. Vuletic, and
  M. D. Lukin, Quantum Kibble-Zurek mechanism and critical dynamics on
  a programmable Rydberg simulator, Nature {\bf 568}, 207 (2019).


\bibitem{PV-02} A. Pelissetto and E. Vicari, Critical phenomena and
  renormalization group theory, Phys. Rep. {\bf 368}, 549 (2002).

\bibitem{GZ-98} R. Guida and J. Zinn-Justin, Critical exponents of the
  N-vector model, J. Phys. A {\bf 31}, 8103 (1998).


\bibitem{CPRV-02} M. Campostrini, A. Pelissetto, P. Rossi, and
  E. Vicari, 25th-order high-temperature expansion results for
  three-dimensional Ising-like systems on the simple-cubic lattice
  Phys. Rev. E {\bf 65}, 066127 (2002).

\bibitem{Hasenbusch-10} M. Hasenbusch, A finite size scaling study of
  lattice models in the three-dimensional Ising universality class,
  Phys. Rev. B {\bf 82}, 174433 (2010).

  
\bibitem{KPSV-16} F. Kos, D. Poland, D. Simmons-Duffin, and A. Vichi,
  Precision islands in the Ising and O($N$) models, J. High Energy Phys.
  {\bf 08}, 036 (2016).

\bibitem{KP-17} M. V. Kompaniets and E. Panzer, Minimally subtracted
  six-loop renormalization of O($n$)-symmetric $\varphi^4$ theory and
  critical exponents, Phys. Rev. D {\bf 96}, 036016 (2017).

\bibitem{Hasenbusch-21} M. Hasenbusch, Restoring isotropy in a
  three-dimensional lattice model: The Ising universality class,
  Phys. Rev. B {\bf 104}, 014426 (2021).
  

  
  \bibitem{Pfeuty-70}
  P. Pfeuty, The one-dimensional Ising model with a transverse field,
  Ann. Phys. {\bf 57}, 79 (1970).

\bibitem{CHPV-02} M. Caselle, M. Hasenbusch, A. Pelissetto, and
  E. Vicari, Irrelevant operators in the two-dimensional Ising model,
  J. Phys. A {\bf 35}, 4861 (2002).

\bibitem{CCCPV-00}
  P. Calabrese, M. Caselle, A. Celi, A. Pelissetto, and E. Vicari, 
  Nonanalyticity of the Callan-Symanzik $\beta$-function of
  two-dimensional O($N$) models, J. Phys. A {\bf 33}, 8155 (2000).

\bibitem{CH-00} M. Caselle and M. Hasenbusch, Critical amplitudes and
  mass spectrum of the 2D Ising model in a magnetic field,
  Nucl. Phys. B {\bf 579}, 667 (2000).

\bibitem{CPRV-98} M. Campostrini, A. Pelissetto, P. Rossi, and
  E. Vicari, Two-point correlation function of three-dimensional
  O($N$) models: The critical limit and anisotropy, Phys. Rev. E
  {\bf 57}, 184 (1998).

  

\bibitem{Schiff-book} L. I. Schiff, Quantum Mechanics (McGraw-Hill
  Kogakusha, Tokyo, 1977).

\bibitem{ROP-08} G. Rigolin, G. Ortiz, and V. H. Ponce, Beyond the
  quantum adiabatic approximation: Adiabatic perturbation theory,
  Phys. Rev. A {\bf 78}, 052508 (2008).
  

  
\bibitem{DP-10} C. De Grandi and A. Polkovnikov, Quantum Quenching,
  Annealing and Computation, edited by A. Das, A. Chandra and
  B. K. Chakrabarti, Lecture Notes in Physics Vol. 802 (Springer,
  Heidelberg, 2010).

  
  \bibitem{CZ-12}
J. I. Cirac and P. Zoller,
Goals and opportunities in quantum simulation,
Nat. Phys. {\bf 8}, 264 (2012).

\bibitem{BDN-12}
  I. Bloch, J. Dalibard, and S. Nascimbe\'ene, Quantum simulations with
  ultracold quantum gases,  Nat. Phys. {\bf 8}, 267 (2012).

\bibitem{BR-12}
  R. Blatt and C. F. Roos, Quantum simulations with trapped ions,
  Nat. Phys. {\bf 8}, 277 (2012).

\bibitem{AW-12}
  A. Aspuru-Guzik and P. Walther, Photonic quantum simulators,
  Nat. Phys. {\bf 8}, 285 (2012).

\bibitem{HTK-12} A. A. Houck, H. E. T\"ureci, and J. Koch,
  On-chip quantum simulation with superconducting circuits,
  Nat. Phys. {\bf 8}, 292 (2012).

  \bibitem{GAN-14}
  I. M. Georgescu, S. Ashhab, and F. Nori,
  Quantum simulation, Rev. Mod. Phys. {\bf 86}, 153 (2014).

    
  
\end{thebibliography}
\end{document}